\documentclass[graybox]{svmult}

\usepackage{amssymb}
\usepackage{amsmath}
\usepackage{epstopdf}
\usepackage{verbatim}
\usepackage{latexsym}
\usepackage{theorem}
\usepackage{fixltx2e}
\usepackage{lscape}
\usepackage{natbib}

\usepackage{mathptmx}       
\usepackage{helvet}         
\usepackage{courier}        
\usepackage{type1cm}        
\usepackage{makeidx}         
\usepackage{graphicx}        
\usepackage{multicol}        
\usepackage[bottom]{footmisc}

\makeindex             

\newcommand\mnras{Mon. Not. R. Astron. Soc.}%
\newcommand\aaps{Astron. Astrophys. Suppl.}%
\newcommand\nat{Nature}
\newcommand\aap{Astron. Astrophys.}%
\newcommand\apj{Astrophys.~J.}%
\newcommand\apjl{Astrophys. J. Lett.}%
\newcommand\apjs{Astrophys. J. Suppl. Ser.}
\newcommand\araa{ARA\&A}
\newcommand\apss{Astrophys. Space Sci.}

\begin{document}

\title*{Evolution and nucleosynthesis of Very Massive Stars}
\author{Raphael Hirschi}
\institute{Raphael Hirschi$^{1,2}$ \at $^{1}$Astrophysics, Lennard-Jones Labs 2.09, EPSAM, Keele University, ST5 5BG, Staffordshire, UK\\
$^{2}$Kavli Institute for the Physics and Mathematics of the Universe (WPI),
University of Tokyo, 5-1-5 Kashiwanoha, Kashiwa, 277-8583, Japan\\
 \email{r.hirschi@keele.ac.uk}}
%
%
\maketitle

\abstract{In this chapter, after a brief introduction and overview of stellar evolution, we discuss the evolution and nucleosynthesis of very massive stars (VMS: $M>100\,M_\odot$) in the context of recent stellar evolution model calculations. This chapter covers the following aspects: general properties, evolution of surface properties, late central evolution, and nucleosynthesis including their dependence on metallicity, mass loss and rotation.
Since very massive stars have very large convective cores during the main-sequence phase, their evolution is not so much affected by rotational mixing, but more by mass loss through stellar winds. Their evolution is never far from a homogeneous evolution even without rotational mixing. All VMS at metallicities close to solar end their life as WC(-WO) type Wolf-Rayet stars.  Due to very important mass loss through stellar winds, these stars may have luminosities during the advanced phases of their evolution similar to stars with initial masses between 60 and 120\,$M_\odot$. A distinctive feature which may be used to disentangle Wolf-Rayet stars originating from VMS from those originating from lower initial masses is the enhanced abundances of neon and magnesium at the surface of WC stars. At solar metallicity, mass loss is so strong that even if a star is born with several hundred solar masses, it will end its life with less than 50\,$M_\odot$ (using current mass loss prescriptions). At the metallicity of the LMC and lower, on the other hand, mass loss is weaker and might enable star to undergo pair-instability supernovae.
}



\section{Introduction}
For a long time, the evolution of VMS
was considered only in the framework of
Pop III stars. Indeed, it was expected that, only in
metal free environments, could such massive stars
be formed, since the absence of dust, an efficient
cooling agent, would prevent a strong
fragmentation of the proto-stellar cloud \citep{BV99, Abel02}\footnote{Note, however, that recent star formation simulations find lower-mass stars forming in groups, similarly to present-day star formation \citep{Stacy10, Greif10}}. 
It came therefore as a surprise when it was
discovered that the most metal-poor low-mass
stars, likely formed from a mixture between the
ejecta of these Pop III stars and pristine
interstellar medium, did not show any signature
of the peculiar nucleosynthesis of the VMS \citep{HEGER02,UN02,CH02, 2005Natur}. While
such observations cannot rule out the existence of
these VMS in Pop III generations (their
nucleosynthetic signature may have been erased by
the more important impact of stars in other mass
ranges), it seriously questions the
importance of such object for understanding the
early chemical evolution of galaxies. Ironically, when the importance of VMS in the context of the first
stellar generations fades, they appear as
potentially interesting objects in the framework of
present day stellar populations. 

For a long time, observations favored a present-day upper mass limit for stars
around 150 $M_\odot$ \citep{FI05,OC05}.
Recently, however, \citet{PAC10} have re-assessed the properties of the brightest
members of the R136a star cluster, revealing exceptionally high
luminosities (see chapter 1 for more details). The comparison between main sequence evolutionary models for rotating and non-rotating stars and observed spectra resulted in high current ($\leq$ 265
$M_{\odot}$) and initial ($\leq  $ 320 $M_{\odot}$) masses for
these stars. The formation scenarios for these VMS are presented in chapter 2.

%
%

The above observations triggered a new interest in the evolution of very massive stars. However, since VMS are so rare, only a few of them are known and we have to rely on stellar evolution models in order to study their properties and evolution. In this chapter, the evolution of VMS will be discussed based on stellar evolution models
calculated using the Geneva stellar evolution code \citep{GVAcode} including the modifications implemented to follow the advanced stages as described in \citet{psn04}. Models at solar ($Z=0.014$), Large Magellanic Cloud (LMC, $Z=0.006$) and Small Magellanic Cloud (SMC, $Z=0.002$) metallicities will be presented \citep[see][for full details about these models]{Liza13}. These models will also be compared to models of normal massive stars calculated with the same input physics at solar
metallicity ($Z=0.014$) presented in \citet{SE12} and \citet{wr12} as well as their extension to lower metallicities \citep{GEE13}. 

In Sect.~2, we review the basics of stellar evolution models and their key physical ingredients. The general properties and early evolution of VMS are presented in Sect.~3. The Wolf-Rayet stars originating from VMS are discussed in Sect.~4. The late evolution and possible fates of the VMS is the subject of Sect.~5. The nucleosynthesis and contribution to chemical evolution of galaxies is discussed in Sect.~6. A summary and conclusions are given in Sect.~7.

\section{Stellar evolution models}
The evolution of VMS is similar enough to more common massive stars that the same stellar evolution codes can be used to study their evolution and corresponding nucleosynthesis. Stellar evolution models require a wide range of input physics ranging from nuclear reaction rates to mass loss prescriptions. In this section, we review the basic equations that govern the structure and evolution of stars as well as some of the key input physics with a special emphasis on mass loss, rotation and magnetic fields.

\subsection{Stellar structure equations}
There are four equations describing the evolution of the structure of stars: the mass, momentum and energy conservation equations and the
energy transport equations, which we recall below. 
On top of that, the equations of the evolution of chemical elements
abundances are to be followed. These equations are discussed in the next section.
In the Geneva stellar evolution code \citep[GENEC, see][]{GVAcode}, which we base our presentation on in this section, the problem is treated in one dimension (1D) and the equations of the evolution of chemical elements
abundances are calculated separately from the structure equations, as in the original
version of Kippenhahn and Weigert \citep{KWH67,kip}. 
In GENEC, rotation is included and spherical symmetry is no longer assumed. The
effective gravity (sum of the centrifugal force and gravity) can in fact no longer be 
derived from a potential and the case is said to be non--conservative.
The problem can still be treated in 1D by assuming that the angular
velocity is constant on isobars. This assumes that there is a strong
horizontal (along isobars) turbulence which enforces constant angular
velocity on isobars \citep{Za92}. The case is referred to as
 ``shellular'' rotation and using reasonable simplifications described in \citet{ROTI}, the usual set of four structure equations (as used for non-rotating stellar models) can be recovered:

\begin{itemize}
\item Energy conservation:
\begin{equation}\label{EQS}
{\partial L_P \over \partial M_P}=\epsilon_{nucl}-\epsilon_{\nu}+
\epsilon_{grav}
=\epsilon_{nucl}-\epsilon_{\nu} -c_{{P}}\frac{\partial \overline T}{\partial t} + \frac{\delta}{\overline \rho}\frac{\partial P}{\partial t}
\end{equation} 
Where $L_P$ is the luminosity, $M_P$ the Lagrangian mass coordinate, and $\epsilon_{nucl}$, $\epsilon_{\nu}$, and $\epsilon_{grav}$ are the energy generation rates per unit mass
for nuclear reactions, neutrinos and gravitational energy changes due to contraction or expansion, respectively. $T$ is the temperature, $c_P$ the specific heat at constant pressure, $t$ the time, $P$ the pressure, $\rho$ the density and $\delta = -\partial \textrm{ln}\rho/\partial \textrm{ln}T$.

\item Momentum equation:
\begin{equation}\label{EQP}
{\partial P \over \partial M_P}=
-{GM_P \over 4\pi r_P^4} f_P 
\end{equation} 
Where $r_P$ is the radius of the shell enclosing mass $M_P$ and $G$ the gravitational constant.

\item Mass conservation (continuity equation):
\begin{equation}\label{EQR}
{\partial r_P \over \partial M_P}={1 \over 4\pi r_P^2 \overline\rho}
\end{equation} 

\item Energy transport equation:
\begin{equation}\label{EQT}
{\partial \ln \overline T \over \partial M_P}  
=-{GM_P \over 4\pi r_P^4 P} f_P {\rm min}[\nabla_{\rm ad}, \nabla_{\rm rad}
{f_T \over f_P}]
\end{equation}
\end{itemize} 
 
\noindent where

\begin{eqnarray*}
\nabla_{\rm{ad}} &=&\left({\partial \ln \overline T \over \partial \ln P}\right)_{\rm{ad}} = \frac{P \delta}{\overline T \overline\rho c_{P}} 
\rm{\ \ (convective\ zones)}, \\
\nabla_{\rm{rad}} &=&\frac{3}{64\pi \sigma G}\frac{\kappa L_P P}{M_P \overline T^4} 
\rm{\ \ (radiative\ zones)},\\
\end{eqnarray*}
where $\kappa$ is the 
total
opacity and $\sigma$ is the Stefan-Boltzmann constant.
\begin{eqnarray*}
f_P&=&{4\pi r_P^4 \over GM_P S_P} {1 \over <g^{-1}>},\\
f_T&=&\left( {4\pi r_P^2 \over S_P}\right)^2 {1 \over <g> <g^{-1}>},\\
\end{eqnarray*}

\noindent $< x >$ is $x$ averaged on an isobaric surface, $\overline x$ is $x$
averaged in the volume separating two successive isobars and the index $P$ 
refers to the isobar with a pressure equal to $P$. $g$ is the effective gravity and $S_P$ is the surface of the isobar \citep[see][ for more
details]{ROTI}. The implementation of the structure equations into other stellar evolution codes are presented for example in \citet{MESA} and \citet{CLS98}.

\subsection{Mass loss}\label{dMdt}

Mass loss strongly affects the evolution of very massive stars as we shall describe below. Mass loss is already discussed in chapter 3 but here we will recall the different mass loss prescriptions used in stellar evolution calculations and how they relate to each other. In the models presented in this chapter, the following prescriptions were used. For main-sequence stars, the prescription for radiative line driven winds from \citet{VN01} was used, which compare rather well with observations \citep{PAC10,muijres11}. 
For stars in a domain not covered by the Vink et~al prescription, the \citet{deJager88} prescription was applied to models with $\log (T_\text{eff}) > 3.7$. For $\log (T_\text{eff}) \leq 3.7$, a linear fit to the data from \citet{sylvester98} and \citet{vloon99} \citep[see][]{Crowther01}  was performed. The formula used is given in Eq.\,2.1 in \citet{BHP12}.

In the stellar evolution simulations, the stellar wind is not simulated self-consistently and a criterion is used to determine when a star becomes a WR star. Usually, a star becomes a WR when the surface hydrogen mass fraction, $X_s$, becomes inferior to 0.3 (sometimes when it is inferior ot 0.4) and the effective temperature, $\log(T_\text{eff})$, is greater than 4.0. 
The mass loss rate used during the WR phase depends on the WR sub-type. For the eWNL phase (when $0.3>X_s>0.05$), the 
\citet{GH08} recipe was used (in the validity domain of this prescription, which usually covers most of the eWNL phase). In many cases, the WR mass-loss rate of \citet{GH08} is lower than the rate of \citet{VN01}, in which case, the latter was used. For the eWNE phase -- when $0.05>X_s$ and the ratio of the mass fractions of ($^{12}$C$+\,^{16}$O)/$^4$He$< 0.03$ -- and WC/WO phases -- when ($^{12}$C$+\,^{16}$O)/$^4$He$ > 0.03$-- the corresponding prescriptions of \citet{NL00} were used. 
Note also that both the \citet{NL00} and \citet{GH08} mass-loss rates account for clumping effects \citep{muijres11}.

As is discussed below, the mass loss rates from \citet{NL00} for the eWNE phase are much larger than in other phases and thus the largest mass loss occurs during this phase. In \citet{PAC10}, the mass loss prescription from \citet{NL00} was used for both the eWNL and eWNE phases (with a clumping factor, $f=0.1$). The models presented in this chapter thus lose less mass than those presented in \citet{PAC10} during the eWNL phase. 

The metallicity dependence of the mass loss rates is commonly included in the following way. The mass loss rate used at a given metallicity, $\dot{M}(Z)$, is the mass loss rate at solar metallicity, $\dot{M}(Z_\odot)$, multiplied by the ratio of the metallicities to the power of $\alpha$: $\dot{M}(Z)= \dot{M}(Z_\odot)(Z/Z_\odot)^\alpha$.
$\alpha$ was set to 0.85 for the O-type phase and WN phase and 0.66 for the WC and WO phases{; and
for WR stars the initial metallicity rather than the actual surface metallicity was used in the equation above following \citet{EV06}}.
$\alpha$ was set to 0.5 for the \citet{deJager88} prescription.

For rotating models, the correction factor described below in Eq.\,\ref{EqMdotRot} is applied to the radiative mass-loss rate.

\subsection{Rotation and magnetic mields}\label{rnb}
The physics of rotation included in stellar evolution codes has been developed extensively
over the last twenty years. A recent review of this development can be found in \citet{MM12}. 
The effects induced by rotation can be divided into three categories. 

1) {\bf Hydrostatic effects}: The centrifugal force changes the 
hydrostatic equilibrium of the star. The star becomes oblate and the equations
describing the stellar structure have to be modified as described above.

2) {\bf Mass loss enhancement and anisotropy}:
Mass loss depends on the opacity and the effective gravity 
(sum of gravity and centrifugal force) at the surface. 
The larger the opacity, the larger the mass loss.
The higher the effective gravity,  
the higher the radiative flux \citep{VZ24} and effective
temperature.
Rotation, via the centrifugal force, reduces the surface effective
gravity at the equator compared to the pole. 
As a result, the radiative flux of the star is larger at the pole than
at the equator. In massive hot stars, since the opacity is dominated by the
temperature--independent electron scattering, rotation enhances mass
loss at the pole. If the opacity increases when the temperature
decreases (in cooler stars), mass loss can be enhanced at the equator 
when the bi-stability is reached (see mass loss chapter for more details).   

For rotating models, the mass loss rates can be obtained by applying a correction factor to the radiative mass loss rate as described in \citet{ROTVI}:
\begin{eqnarray}
\dot{M}(\Omega) &=& F_{\Omega}\cdot \dot{M}(\Omega=0)= F_{\Omega}\cdot \dot{M}_{\rm rad}  \nonumber \\
& &{\rm with} \hspace{.3cm}F_{\Omega}=\frac{(1-\Gamma)^{\frac{1}{\alpha}-1}}{\left[ 1-\frac{\Omega^2}{2\pi G \rho_{\rm m}} - \Gamma \right]^{\frac{1}{\alpha}-1}}
\end{eqnarray}\label{EqMdotRot}
where $\Gamma=L/L_\text{Edd}=\kappa L / (4\pi cGM)$ is the Eddington factor (with $\kappa$ the total opacity), and $\alpha$ the $T_\text{eff}-$dependent force multiplier parameter. Enhancement factors ($F_{\Omega}$) are generally close to one but they may become very large when $\Gamma \gtrsim 0.7$ or $\Omega/\Omega_{\rm crit}>0.9$ \citep[see][for more details]{ROTVI,GMM11}. If critical rotation, where the centrifugal force balances gravity at the equator, is reached, mechanical mass loss may occur and produce a decretion disk \citep[see][for more details]{KMO11}. In most stellar evolution codes, the mass loss is artificially enhanced when $\Omega/\Omega_{\rm crit} \gtrsim 0.95$ to ensure that the ratio does not become larger than unity but multi-dimensional simulations are required to provide new prescriptions to use in stellar evolution codes.

For mass loss rates, $\dot{M}(\Omega=0)$, the following prescriptions are commonly used: \citet{VKL01} for radiatively driven wind of O-type stars, \citet{NuLa00} for Wolf-Rayet stars and \citet{Ja88} for cooler stars not covered by the other two prescriptions and for which dust and pulsation could play a role in the driving of the wind.

3) {\bf Rotation driven instabilities}:
The main rotation driven instabilities are
horizontal turbulence, meridional circulation and dynamical and secular
shear \citep[see][for a comprehensive description of rotation-induced instabilities]{AM09}. 

Horizontal turbulence
corresponds to turbulence along the isobars. If this turbulence is
strong, rotation is constant on isobars and the situation is usually
referred to as ``shellular rotation'' \citep{Za92}. The
horizontal turbulence is expected to be stronger than the vertical
turbulence because there is no restoring buoyancy force along
isobars \citep[see][ for recent development on this topic]{Mh03}.

\begin{figure}
\centering
\includegraphics[width=0.8\textwidth]{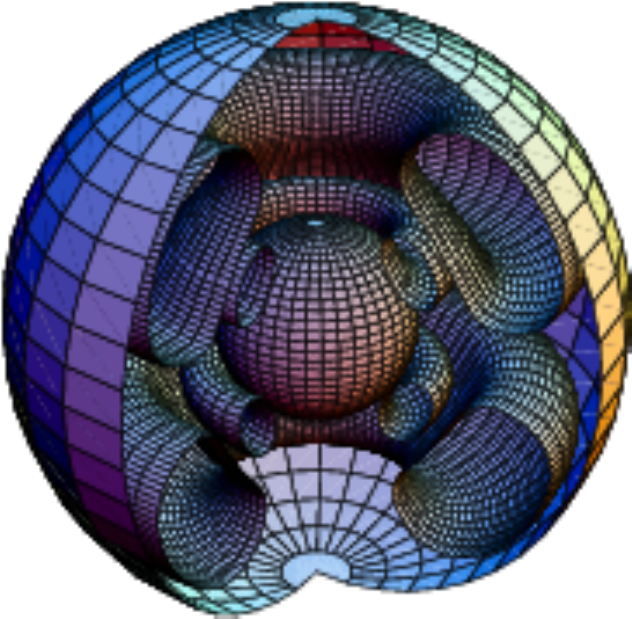}
\caption{Streamlines of meridional circulation in a rotating 20 M$_\odot$
model with  solar metallicity and
$v_{\rm ini}=300$ km\,s$^{-1}$ at the beginning of the H--burning phase. 
The streamlines are in the meridian plane.
In the upper hemisphere on the right section, matter
is turning counterclockwise along the outer streamline and
clockwise along the inner one. The outer sphere is the star surface and has
a radius equal to 5.2 R$_\odot$. The inner sphere is the outer boundary of the
convective core. It has a radius of 1.7 R$_\odot$ \citep[Illustration taken from][]{ROTVIII}.\label{mc}}
\end{figure}

Meridional circulation, also referred to as Eddington--Sweet circulation,
arises from the local breakdown of radiative equilibrium in rotating
stars. This is due to the fact that surfaces of constant temperature do
not coincide with surfaces of constant pressure. Indeed, since rotation
elongates isobars at the equator, the temperature on the same isobar is
lower at the equator than at the pole. This induces large scale
circulation of matter, in which matter usually rises at the pole and 
descends at the equator (see Fig.~\ref{mc}). 

In this situation, angular momentum is
transported inwards. It is however also possible for the circulation to
go in the reverse direction and, in this second case, angular momentum is
transported outwards. Circulation corresponds to an
advective process, which is different from diffusion because the latter 
can only erode gradients. Advection can either
build or erode angular velocity gradients \citep[see][ for more details]{ROTIII}.

Dynamical shear occurs when the excess energy contained in
differentially rotating layers is larger then the work that needs to be
done to overcome the buoyancy force. The criterion for stability against dynamical shear instability is the 
Richardson criterion:

\begin{equation}
Ri=\frac{N^2}{(\partial U/\partial z)^2} > \frac{1}{4}=Ri_c,
\end{equation} where 
$U$ is the horizontal velocity, $z$ the vertical coordinate and $N^2$
the Brunt--V\"ais\"al\"a frequency.

The critical value of the Richardson criterion, $Ri_c=1/4$, corresponds to the situation where the 
excess
kinetic energy contained in the differentially rotating layers
is equal to the work done against
the restoring force of the density gradient (also called buoyancy force). 
It is
therefore used by most authors as the limit for the occurrence of the 
dynamical shear.
However, studies by \citet{CA02} show that turbulence may occur as
long as $Ri \lesssim Ri_c \sim 1$. This critical value is consistent with 
numerical
simulations done by \citet{BH01} where they find shear
mixing for values of $Ri$ greater than $1/4$ (up to about $1.5$).\\

Different dynamical shear diffusion coefficients, $D$, can be found 
in the literature. The one used in GENEC is:

\begin{equation}
D=\frac{1}{3}vl
=\frac{1}{3}\ \frac{v}{l}\ l^2
=\frac{1}{3}\ r\frac{\mathrm{d}\,\Omega}{\mathrm{d}\,r} \ \Delta r^2
=\frac{1}{3}\ r\Delta\Omega\ \Delta r
\label{formds}
\end{equation}where $r$ is the mean radius of the zone where the 
instability occurs,
$\Delta\Omega$ is the variation of $\Omega$ over this zone and
$\Delta r$ is the extent of the zone. The zone is the reunion of consecutive shells
where $Ri< Ri_c$ \citep[see][for more details and references]{psn04a}. 

If the
differential rotation is not strong enough to induce dynamical shear, it
can still induce the secular shear instability when thermal turbulence
reduces the effect of the buoyancy force. 
The secular shear instability occurs therefore on the thermal 
time scale, which is
much longer than the dynamical one. 
Note that the way the inhibiting effect of the 
molecular weight ($\mu$) gradients on secular shear 
is taken into account impacts strongly the efficiency of the shear.
In some work, the inhibiting effect of $\mu$--gradients is so strong that
secular shear is suppressed below a certain threshold value 
 of differential rotation \citep{HLW00}. In other work \citep{ROTII}, 
thermal instabilities and horizontal turbulence reduce the
inhibiting effect of the $\mu$--gradients. As a result, shear is 
not suppressed below a threshold value of differential rotation
but only decreased when $\mu$--gradients are present.

There are other minor instabilities induced by rotation: the GSF
instability \citep{GS67,Fr68,HM10}, the ABCD instability \citep{KS83,HLW00} 
and the Solberg--H\o iland instability \citep{kip}. The GSF instability is induced by axisymmetric
perturbations. The ABCD instability is a kind of horizontal convection.
Finally, Solberg--H\o iland stability criterion is the criterion that
should be used instead of the Ledoux or Schwarzschild criterion in
rotating stars. However, including the dynamical shear instability also takes into account the Solberg--H\o iland instability \citep{psn04a}.

\subsubsection{Transport of angular momentum}

For shellular rotation, the equation of transport of angular
momentum \citep{Za92} in the vertical direction is 
(in lagrangian coordinates):

\begin{equation}
\rho\frac{d}{dt}
\left( r^2 \Omega\right)_{M_{r}} = 
 \frac{1}{5 r^2}  \frac{\partial}{\partial r}
\left(\rho r^4 \Omega U(r)\right)
  + \frac{1}{r^2} \frac{\partial}{\partial r}
\left(\rho D r^4 \frac{\partial\Omega}{\partial r} \right) \;,
\label{c2to}
\end{equation}

\noindent where $\Omega(r)$ is the mean angular velocity at level r,
$U(r)$  the vertical component of the meridional circulation
velocity and $D$  the  diffusion coefficient due to the sum
of the various turbulent diffusion processes (convection, shears and other rotation induced instabilities apart from meridional circulation). Note that angular momentum is
conserved in the case of contraction or expansion.
The first term on the right hand side, corresponding to meridional
circulation, is an \textit{advective} term. The
second term on the right hand side, which corresponds to the diffusion
processes, is a \textit{diffusive} term. The correct treatment of advection is very
costly numerically because Eq. \ref{c2to} is a fourth order equation
\citep[the expression of $U(r)$ contains third 
order derivatives of $\Omega$, see][]{Za92}.
This is why some research groups treat meridional circulation in a
diffusive way \citep[see for example][]{HLW00} with the risk of
transporting angular momentum in the wrong direction (in the case
meridional circulation builds gradients).

\subsubsection{Transport of chemical species}

The transport of chemical elements is also governed by a 
diffusion--advection equation like Eq. \ref{c2to}.
However, if the
horizontal  component of the turbulent diffusion is large,
the vertical advection of the elements (but not that of the
angular momentum) can be treated as  a simple diffusion
\citep{CZ92} with a diffusion coefficient
$D_{\rm eff}$,

\begin{equation}
D_{\rm eff} = \frac{\mid rU(r) \mid^2}{30 D_h} \; ,
\label{c2deff}
\end{equation} 

\noindent where $D_h$ is the coefficient of horizontal
turbulence \citep{Za92}. Equation \ref{c2deff} 
expresses that the vertical
advection of chemical elements is severely inhibited by the
strong horizontal turbulence characterized by $D_h$. 
 The change of the mass fraction $X_i$ of the chemical species 
\textit{i} is simply

\begin{equation}
\left(\frac{dX_i}{dt} \right)_{M_r} =
\left(\frac{\partial  }{\partial M_r} \right)_t
\left[ (4\pi r^2 \rho)^2 D_{\rm mix} \left( \frac{\partial X_i}
{\partial M_r}\right)_t
\right] + \left(\frac{d X_i}{dt} \right)_{\rm nuclear} ,
\label{c2tc}
\end{equation} 

\noindent where the second term on the right accounts for
 composition changes due
to nuclear reactions. The coefficient $D_{\rm mix}$ is the sum 
$D_{\rm mix} = D +D_{\rm eff}$,
where D is the term  appearing in Eq. \ref{c2to} and 
$D_{\rm eff}$ accounts
for the combined effect of advection and horizontal 
turbulence.

\subsubsection{Interaction between rotation and magnetic fields}

Circular spectro-polarimetric surveys have obtained evidence for the presence of magnetic field at the surface 
of OB stars \citep[see e.g. the review by ][and references therein]{WFM11}. 
The origin of these magnetic fields is still unknown. It might be fossil fields 
or fields produced through a dynamo mechanism.

The central question for the evolution of massive stars is whether a dynamo is at work in internal radiative zones. This could have far reaching consequences concerning the mixing of the elements and the loss of angular momentum.
In particular, the interaction between rotation and magnetic fields in the stellar interior strongly affects the angular momentum retained in the core and thus the initial rotation rate of pulsars and which massive stars could die as long \& soft gamma-ray bursts (GRBs), see \citet{VGH11} and the discussion in Sect.~6 in \citet[][and references therein]{wr12}. 

The interplay between rotation and magnetic field has been studied in stellar evolution calculations using the Tayler--Spruit  dynamo \citep{Sp02, ROTBIII}. Some numerical simulations confirm the existence of a magnetic instability, however the existence of the dynamo is still controversial \citep{Br06,ZBM07}.

The Tayler-Spruit dynamo is based on the fact that a purely toroidal field $B_{\varphi}(r, \vartheta)$, even very weak, in a stable stratified  star is unstable on an Alfv\'en timescale $1/\omega_{\mathrm{A}}$. This is the first magnetic instability to appear. It is non--axisymmetric of  type  $m=1$ \citep{Sp02}, occurs under a wide range of conditions and is characterized by a low threshold and a  short growth time.  In a rotating star, the instability is also present, however the  growth rate $\sigma_{\mathrm{B}}$ of the instability is, if $\omega_{\mathrm{A}} \ll \Omega$,
\begin{eqnarray}
\sigma_{\mathrm{B}} \, = \, \frac{\omega_{\mathrm{A}}^2}{\Omega}  \; ,
\label{sigcoriolis}
\end{eqnarray}
\noindent
instead of the Alfv\'en frequency $\omega_{\mathrm{A}}$, because the growth rate of the instability is reduced by the Coriolis force \citep{Sp02}. One usually has the following ordering of the different frequencies, 
$
N \, \gg  \, \Omega \, \gg \, \omega_{\mathrm{A}} .$
In the Sun, one has $N \approx  10^{-3}$ s$^{-1}$, $\Omega = 3 \times 10^{-6}$ s$^{-1}$ and 
 a field of 1 kG would give  an Alfv\'en frequency as low as $\omega_{\mathrm{A}} = 4 \times 10^{-9}$
 s$^{-1}$ (where $N^2$ is the Brunt--V\"ais\"al\"a frequency).

This theory enables us to establish the two quantities that we are mainly interested in for stellar evolution: the magnetic viscosity $\nu$, which expresses the mechanical coupling due to the  magnetic field $\vec{B}$, and the magnetic diffusivity $\eta$, which expresses the transport by a magnetic instability and thus also the damping of the instability. 
The parameter $\eta$ also expresses the
vertical transport of the chemical elements and enters Eq.~\ref{c2tc}, while the viscosity
$\nu$ determines the vertical transport of the angular momentum
by the magnetic field and enters the second term on the right-hand side of Eq.~\ref{c2to}.

\begin{figure}
\centering
\begin{tabular}{cc}
\includegraphics[width=0.4\textwidth,clip=]{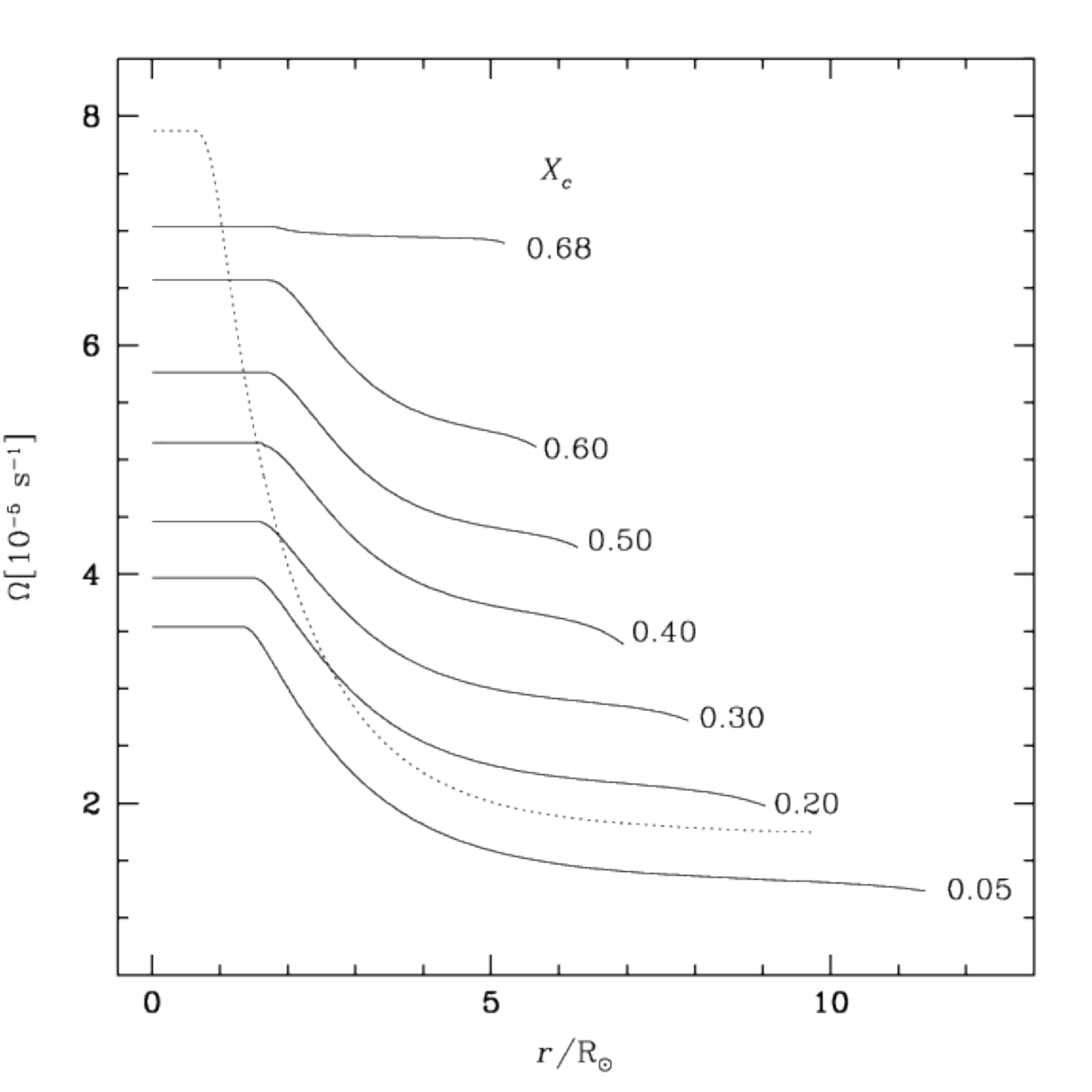} &
\includegraphics[width=0.4\textwidth,clip=]{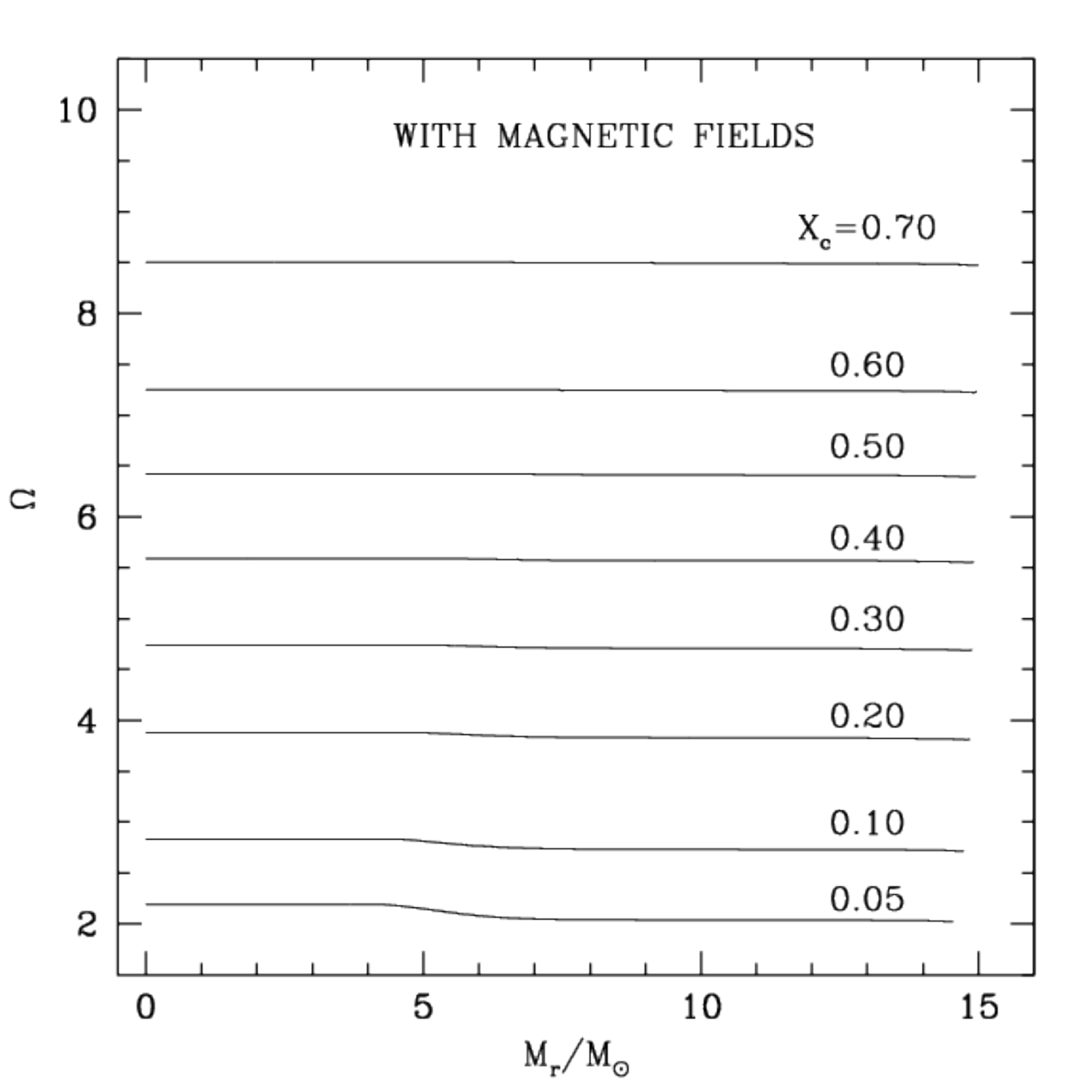} 
\end{tabular}
\caption{Left: evolution of the angular velocity $\Omega$ 
 as a function of the distance to the center
in a 20 M$_\odot$ star with $v_{\rm ini}$ = 300 km s$^{-1}$. 
$X_c$ is the hydrogen mass fraction at the center.
The dotted line shows the profile when the He--core contracts at the end
of the H--burning phase. Right: rotation profiles
at various stages of evolution  (labelled by the central H content $X_{\mathrm{c}}$) of a 15 M$_{\odot}$ 
model with $X=0.705, Z=0.02$, an initial velocity of 300 km s$^{-1}$ and magnetic field from the Tayler--Spruit dynamo
\citep[taken from][]{ROTBIII}.}
\label{Omprofile}
\end{figure}

Figure \ref{Omprofile} shows the differences in the internal $\Omega$--profiles
during the evolution of a 20 M$_{\odot}$ star with and without magnetic field created by the Tayler--Spruit dynamo.  Without magnetic field, the star has a significant differential rotation, while  $\Omega$ is almost constant when a  magnetic field created by the dynamo is present. It is not perfectly constant,
otherwise there would be no dynamo. In fact, the rotation rapidly adjusts itself
to the minimum differential rotation necessary to sustain the dynamo.
One could then assume that the mixing of chemical elements is suppressed by magnetic fields. This is, however, not the case since the interplay between magnetic fields and the meridional circulation tend to lead to more mixing in models including magnetic fields compared to models not including magnetic fields \citep{ROTBIII}. Fast rotating models of GRB progenitors calculated by \citet{YLN06} also experience a strong chemical internal mixing leading to the stars undergoing quasi-chemical homogeneous evolution. The study of the interaction between rotation and magnetic fields is still under development \citep[see e.\,g.][for a different rotation-magnetic field interaction theory, the $\alpha-\Omega$ dynamo, and its impact on massive star evolution]{PCT12} and the next ten years will certainly provide new insights on this important topic.

\subsubsection{Other input physics}
The other key input physics that are essentials for the computation of stellar evolution models are: nuclear reactions, mass loss prescriptions (discussed above), the equation of state, opacities and neutrino losses. Stellar evolution codes are now able to include larger and more flexible nuclear reaction network \citep[see e.~g.][for a description of the implementation of a flexible network in GENEC]{boron10}. Nuclear physics and other inputs are described for other codes for example in \citet{MESA,CLS98}.

In this chapter, the evolution of single stars is discussed. We refer the reader to \citet{L12} for a review of the impact of binarity on massive star evolution and to \citet{SIM14} for the possible impact of a binary companion on the properties of VMS. Note that the mass transfer efficiency prescriptions used in binary model represent an important uncertainty, especially for VMS with high luminosities.

\section{General properties and early evolution of VMS}
\subsection{VMS evolve nearly homogeneously}

\begin{figure}
\centering
\includegraphics[width=0.8\textwidth,clip=]{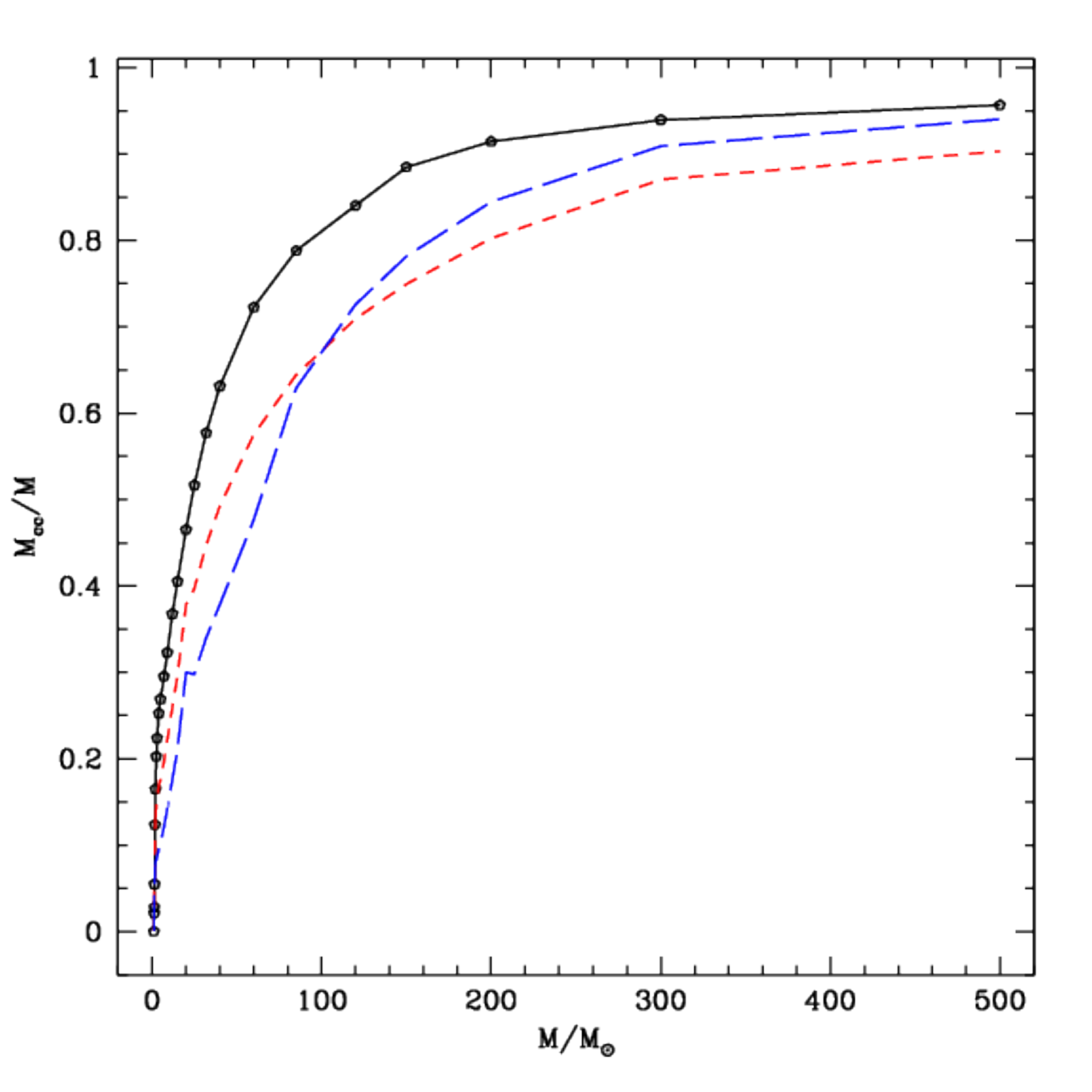}
\caption{Mass fraction of the convective core in non-rotating
solar metallicity models. This figure and all the following figures are taken from \citet{Liza13}. Models with initial masses superior
or equal to 150 $M_\odot$ are from \citet{Liza13}. Models
for lower initial masses are from Ekstr\"om et al. (2012).
The continuous line corresponds to the ZAMS, the short-dashed line to
models when the mass fraction of hydrogen at the centre, $X_c$,  is 0.35, and
the long-dashed line to models when $X_c$ is equal to 0.05.
}\label{fig:ccom}
\end{figure}

\begin{figure*}
\centering
\begin{tabular}{cc}
\includegraphics[width=0.4\textwidth,clip=]{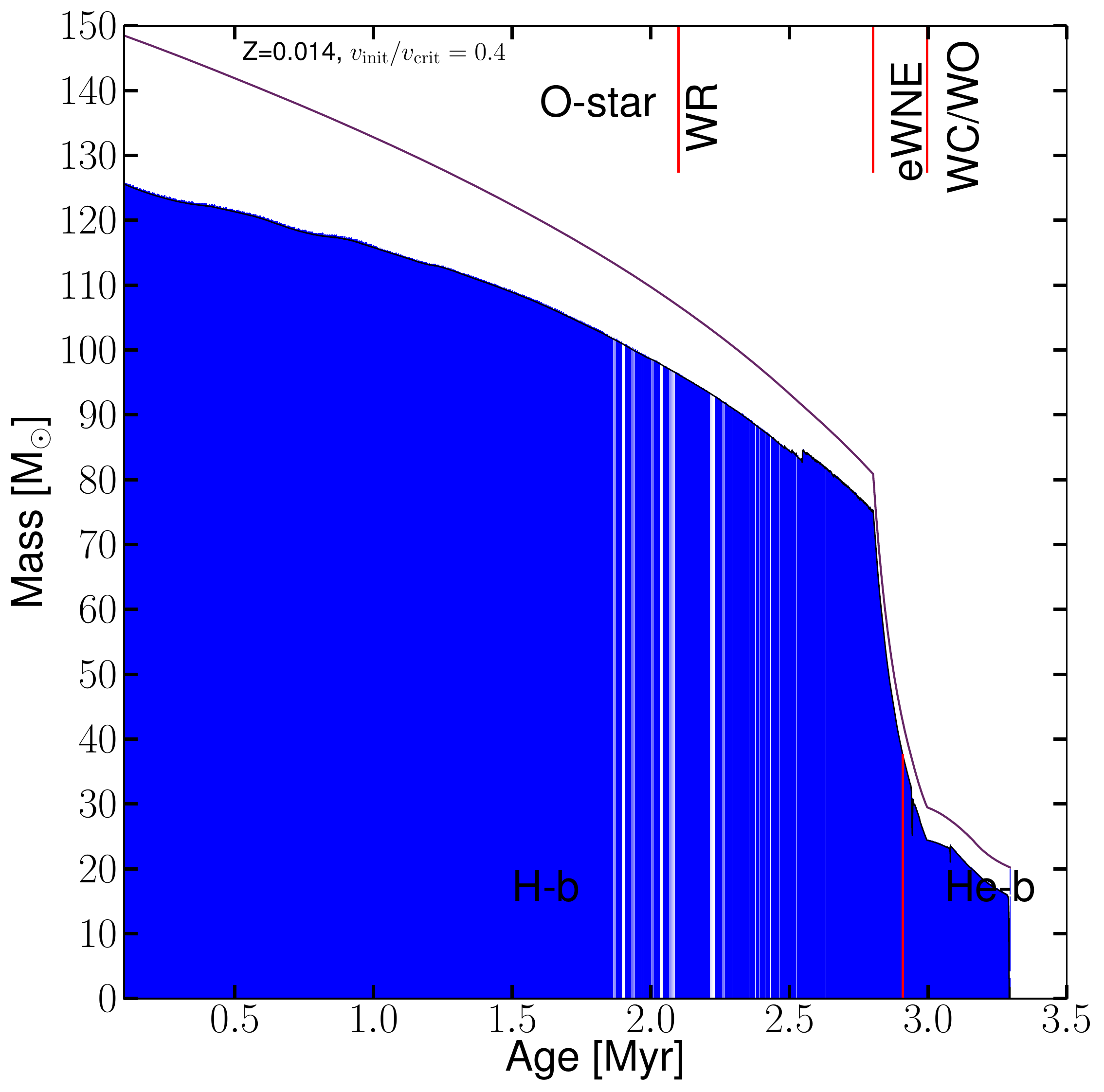} &
\includegraphics[width=0.4\textwidth,clip=]{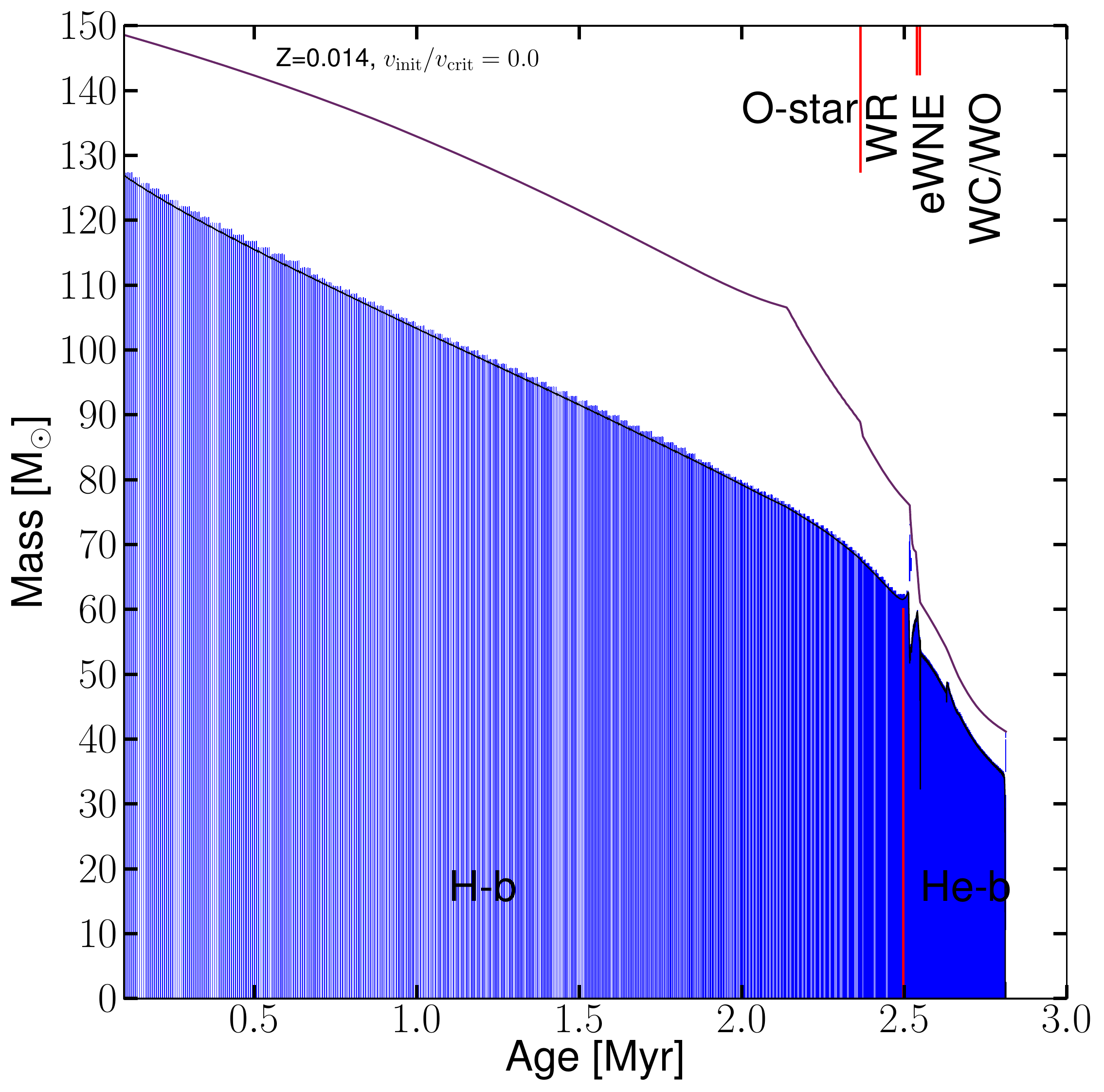} \\
\includegraphics[width=0.4\textwidth,clip=]{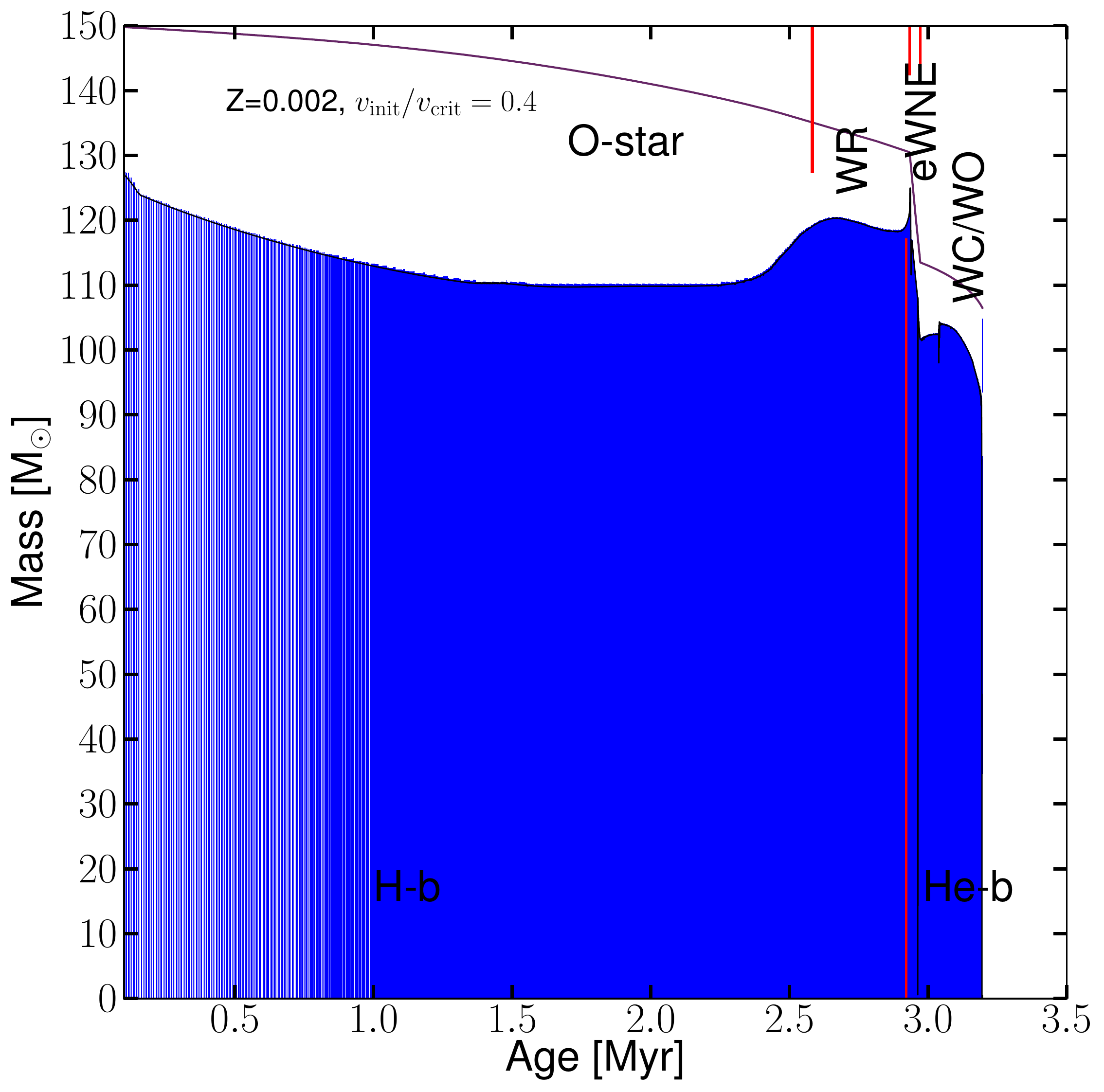} &
\includegraphics[width=0.4\textwidth,clip=]{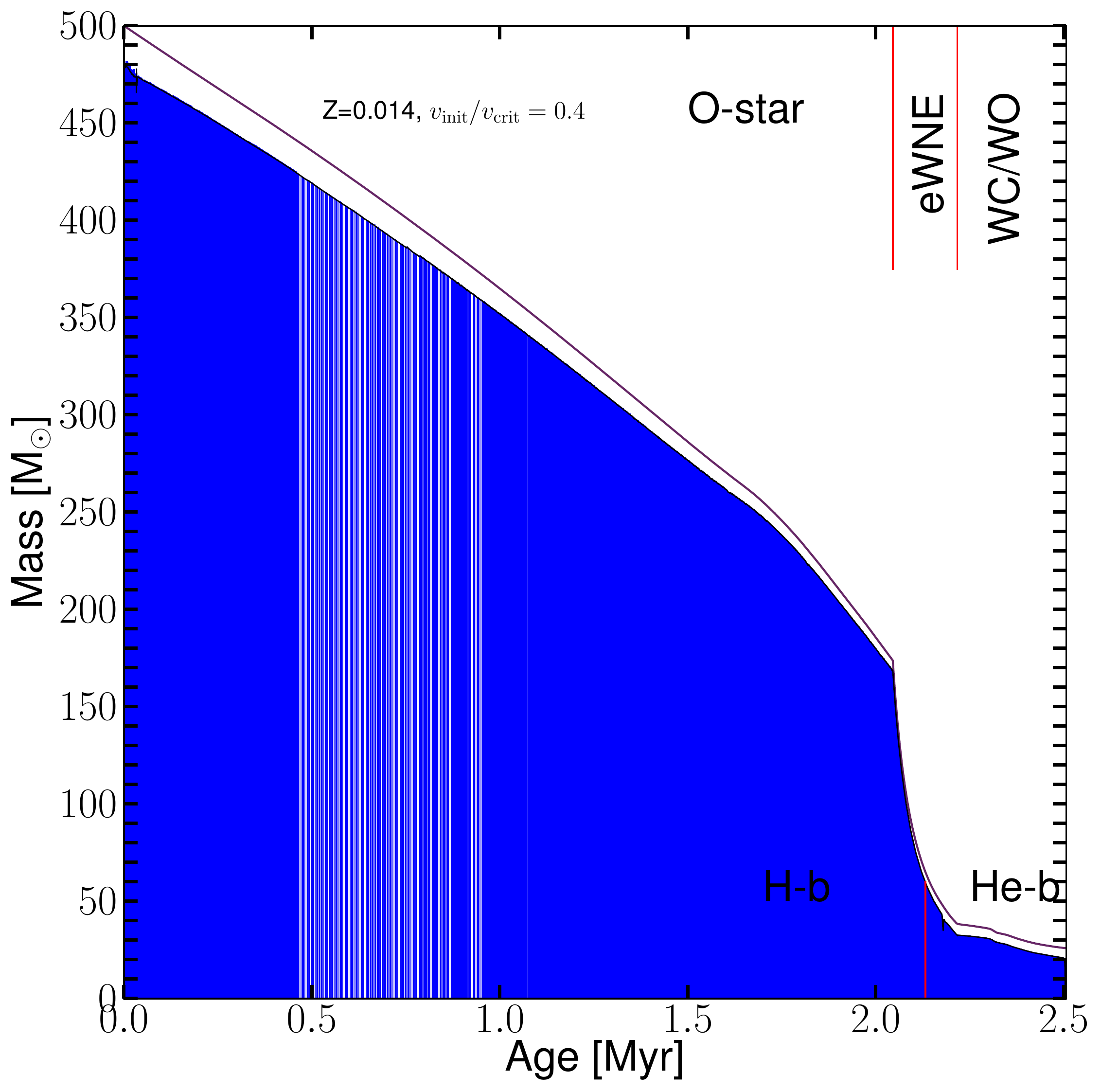} 
\end{tabular}
\caption{Structure evolution as a function of age for selected models: solar metallicity 150 $M_\odot$ rotating ({\it top-left}) and non-rotating ({\it top-right}) models, rotating SMC metallicity 150 $M_\odot$ model ({\it bottom-left}) and rotating solar metallicity 500 $M_\odot$ model ({\it bottom-right}). The blue zones represent the convective regions. The top solid black line indicates the total mass of the star and vertical red markers are given for the different phases (O-type, WR=eWNL, eWNE and WC/WO) at the top of the plots. The transition between H- and He-burning phases is indicated by the red vertical line at the bottom of the plots.}\label{fig:kip_age}
\end{figure*}

Probably the main characteristic that makes VMS quite different from their lower mass siblings is the fact that
they possess very large convective cores during the MS phase. They therefore evolve quasi-chemically homogeneously even if there is no mixing (due e.\,g. by rotation) in radiative zones as discussed in \citet{M80}.
To illustrate this last point, Fig.~\ref{fig:ccom} shows the convective core mass fraction for 
non-rotating massive stars at solar metallicity.
It is apparent that the convective cores for masses above 150 $M_\odot$
extend over more than 75\% of the total mass of the star.

Figure~\ref{fig:kip_age} shows how age, metallicity and rotation influence this mass fraction. 
Comparing the {\it top-left} and {\it bottom-left} panels showing the rotating 150 $M_\odot$ models at solar and SMC metallicities ($Z$), respectively, we can see that the convective core occupies a { very slightly larger fraction of the total mass at SMC metallicity on the ZAMS. As for lower-mass massive stars, this is due to a lower CNO content leading to higher central temperature. This effect is counterbalanced by the lower opacity (especially at very low metallicities) and the net change in convective core size is small. As the evolution proceeds mass loss is weaker at lower $Z$ and thus the total mass decreases slower than the convective core mass. This generally leads to a smaller fraction of the total mass occupied by the convective core in the SMC models.}

We can see the impact of rotation by comparing the rotating ({\it top-left}) and non-rotating ({\it top-right}) 150 $M_\odot$ models. The convective core size remains higher in the rotating model due to the additional mixing in radiative zones. We can see that rotation induced mixing can even lead to an increase of the convective core size as is the case for the SMC model ({\it bottom-left}). This increase is typical of quasi-chemically homogeneous evolution also found in previous studies \citep[see][and citations therein]{YDL12}.
The rotating 500 $M_\odot$ model ({\it bottom-right} panel) evolves quasi-homogeneously throughout its entire evolution, even with an initial ratio of the velocity to the critical velocity of 0.4. 

These features, very large convective cores and quasi-chemi homogeneous evolution, 
are a key factors governing their evolution as is discussed below.

\subsection{Evolutionary tracks}

\begin{figure}
\center
\includegraphics[width=0.5\textwidth,clip=]{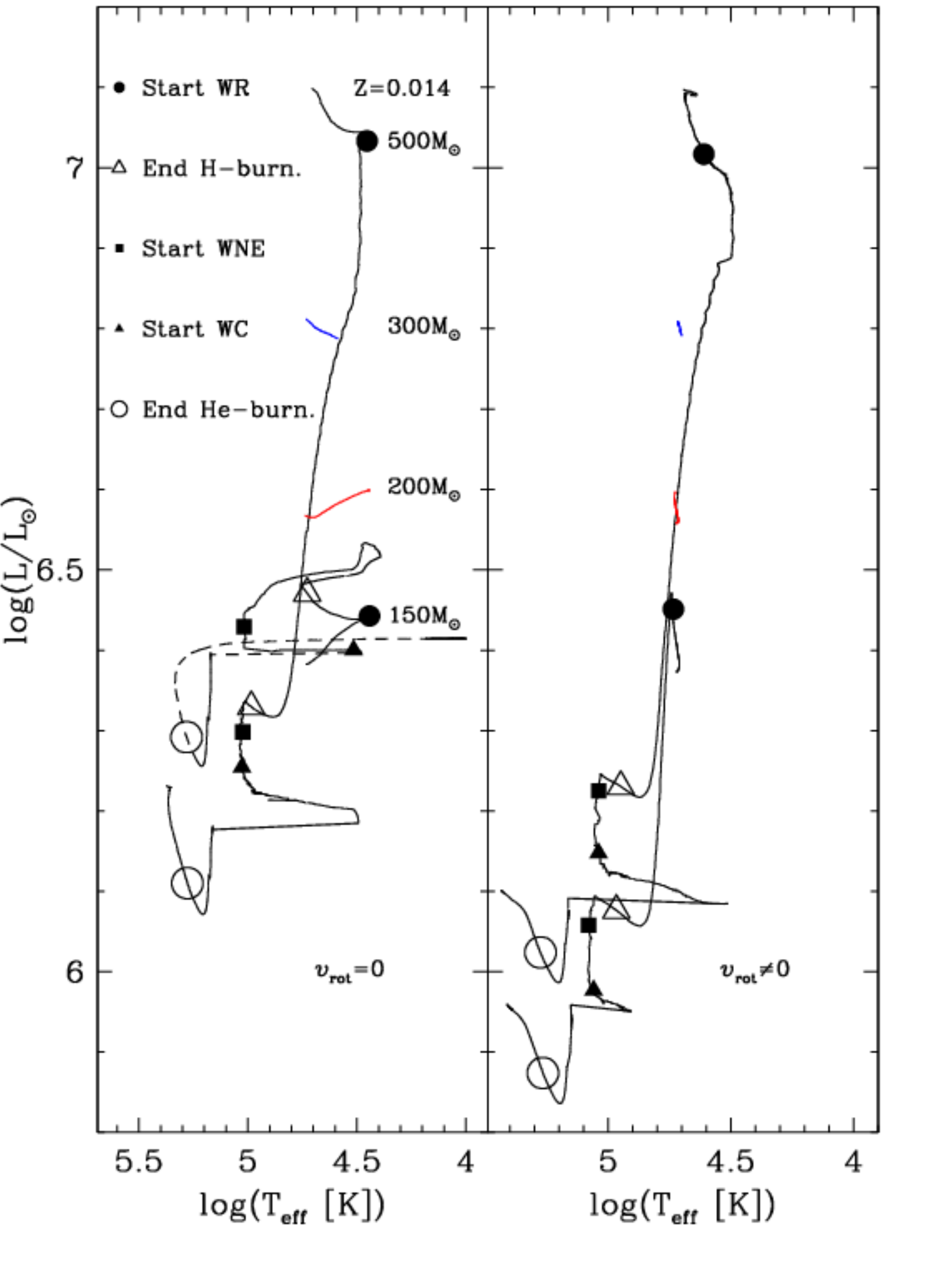}
\caption{HR diagram from 150 up to 500 $M_\odot$ at solar 
metallicity for non-rotating ({\it left}) and rotating ({\it right}) models, respectively. Key stages are indicated along the tracks. Only the first portion (up to start of WR phase) of the tracks for the 200 and 300 $M_\odot$ are shown.}\label{fig:hrdsolar}
\end{figure}

\begin{figure}
\center
\includegraphics[width=0.5\textwidth,clip=]{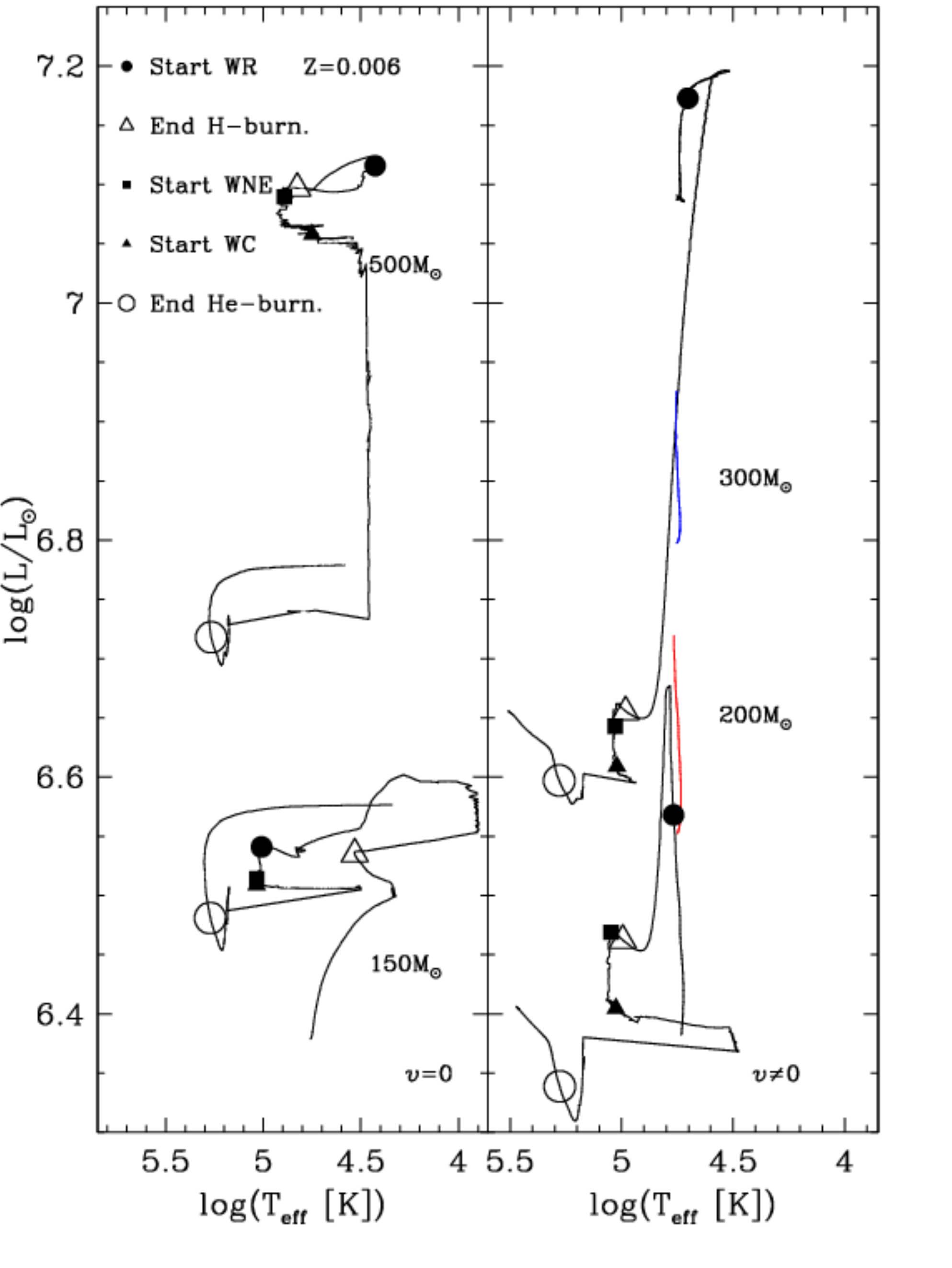}\includegraphics[width=0.5\textwidth,clip=]{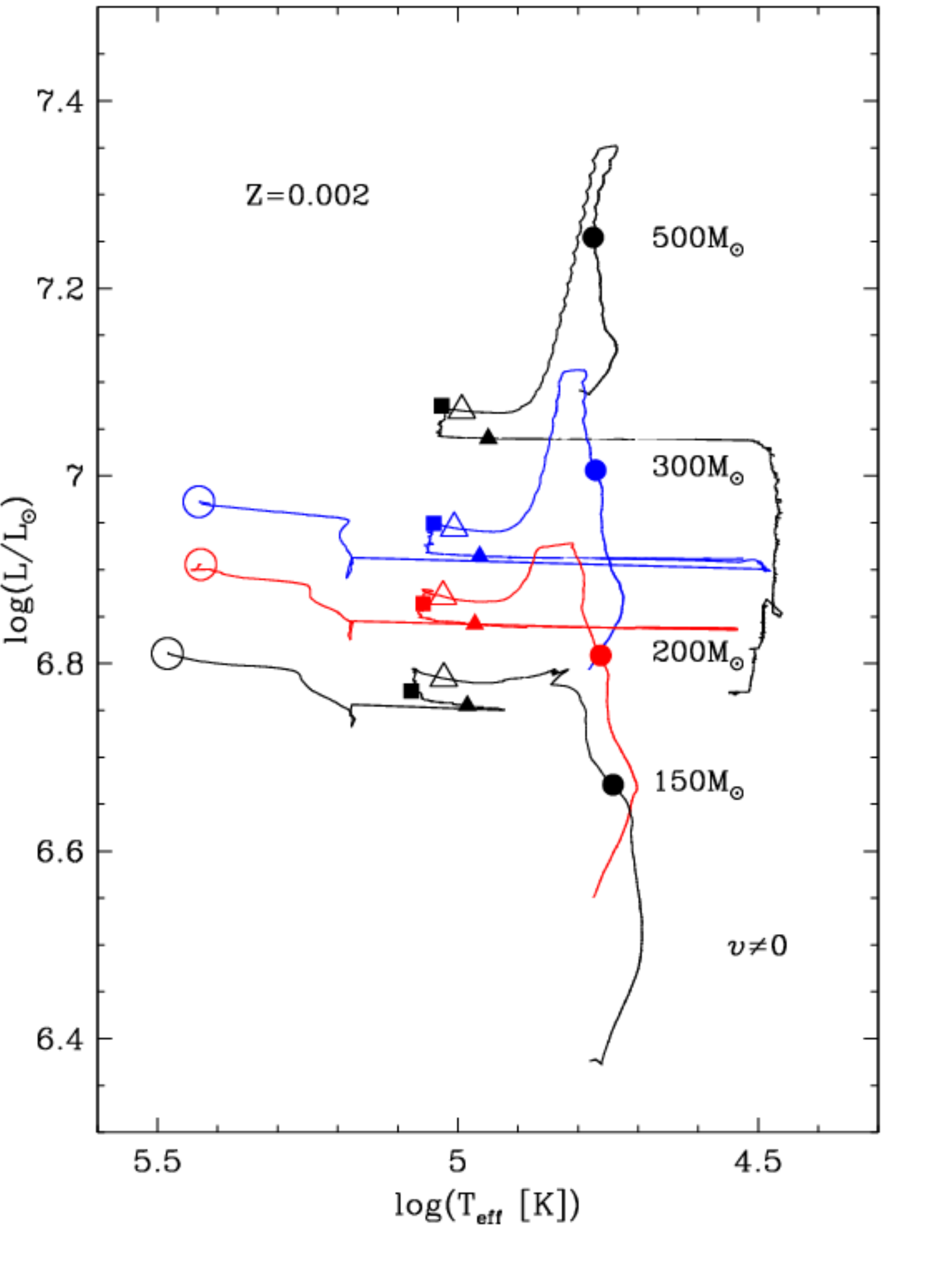}
\caption{Same as Fig.~\ref{fig:hrdsolar} for {\it right}: LMC models ($Z=0.006$)  and {\it right}: SMC rotating models ($Z=0.002$).}
\label{fig:hrdlmcsmc}
\end{figure}

In Figs.~\ref{fig:hrdsolar} and \ref{fig:hrdlmcsmc}, we present the evolutionary tracks of models with initial masses between 150 and 500 $M_\odot$ at various metallicities.
Other properties of VMS models at the end of H- and He- burning stages are given in Tables~\ref{Table:endH} and \ref{Table:endHe}, respectively.


A very first striking feature is that these massive star evolve vertically in the HR diagram (HRD) covering only very restricted ranges in effective temperatures but a 
very large range in luminosities. This is typical of an evolution mainly governed by mass loss and also by a strong internal mixing (here due to convection).

Let us now describe in more details the evolution of the non-rotating 500 $M_\odot$ model at solar metallicity (see Fig.~\ref{fig:hrdsolar}).
In general, the luminosity of stars increases during the MS phase. Here we have that during that phase, 
the luminosity decreases slightly by about 0.1 dex.
This is the consequence of very high mass loss rates (of the order of $7 \times 10^{-5}$ $M_\odot$ per year) already at very early evolutionary stages.

At an age of 1.43  million years, the mass fraction of hydrogen at the surface becomes inferior to 0.3, the star enters into the WR phase and has an actual mass decreased by about 40\%
with respect to its initial value. At that time the mass fraction of hydrogen in the core is 0.24. Thus this star enters the WR phase while still burning hydrogen in its core and having nearly the same amount of hydrogen at the centre and at the surface, illustrating the nearly homogeneous nature of its evolution \citep[see also][]{M80}.  Typically for this model, the convective core encompasses nearly 96\% of the total mass on the ZAMS (see also Fig. \ref{fig:ccom}).

At an age equal to 2.00 Myr, the mass fraction of hydrogen is zero in the core ($X_c=0$). The star has lost a huge amount of mass through stellar winds and has at this stage an actual mass of 55.7 $M_\odot$.  So, since the entrance into the WR phase, the star has lost about 245 $M_\odot$, i.e. about half of its total mass. This strong mass loss episode translates into the HR diagram
by a very important decrease in luminosity. Note that when $X_c$ is zero, the convective core still encompass 80\% of the total stellar mass!

The core helium burning phase last for about 0.3 Myr, that means slightly more than 15\% of the MS lifetime. 
At the end of the core He burning phase, the actual mass of the star is 29.82 $M_\odot$, its age is 2.32 My, the mass fraction of helium at the surface is 0.26. The total WR phase lasts for 0.88 My, that means about 38\% of the total stellar lifetime.

It is interesting to compare the evolution of the 500 $M_\odot$ stellar model with that of the 150 $M_\odot$ model. In contrast to the 500 $M_\odot$ model, the 150 $M_\odot$ increases in luminosity during the MS phase. Looking at the HRD we see that the O-type star phases of the 150 and 500 $M_\odot$ models cover more or less the same effective temperature range. This illustrates the well known fact that the colors of stars for this mass range does not change much with the initial mass. 

When the stars enters into the WR phase, in contrast to the case of the 500 $M_\odot$ where the luminosity decreases steeply, the luminosity of the 150 $M_\odot$ model continues to increase a little. The luminosities of the two models when the hydrogen mass fraction at the surface becomes inferior to 10$^{-5}$ differ by just a little more than 0.1 dex.  The effective temperatures are similar. Thus one expects stars from very different initial masses to occupy similar positions in the HRD (the 500 $M_\odot$ star being slightly less luminous than the 150 $M_\odot$ during the WR phase).
We note that after the end of the core He-burning phase, the star evolves to the red and terminate its lifetime around an effective temperature of Log $T_{\rm eff}$ equal to 4. This comes from the  
core contraction at the end of core He-burning which releases energy and leads to an envelope expansion akin to the expansion of the envelope at the end of the MS \citep[see also][]{YH11}. 

The duration of the core H-burning phase of the 150 $M_\odot$ model is not much different from the one of the 500 $M_\odot$ model being 2.5 My instead of 2 My. 
The core He-burning lifetime lasts for 0.3 My as for the 500 $M_\odot$. 
The total duration of the WR phase is 0.45 My, thus about half of the WR duration for the 500 $M_\odot$.

The 200 $M_\odot$ model has an evolution  similar to the 150 $M_\odot$ model, 
while the 300 $M_\odot$ has an evolution similar to the 500 $M_\odot$.

Let us now consider how rotation changes the picture. The right panel of 
Fig.~\ref{fig:hrdsolar} shows the evolutionary tracks of the $Z=0.014$ rotating models in a similar way to the tracks of the non-rotating models in the left panel. The changes brought by rotation are modest. 
This is expected because of two facts: first, in this high mass range, the evolution is more impacted by mass loss than by rotation, second, stars are already well mixed by the large convective cores.
One notes however a few differences between the non-rotating and rotating models. One of the most striking difference is the fact that the models during their O-type phase evolve nearly vertically when rotation is accounted for. This is the effect of rotational mixing which keeps the star more chemically homogeneous than in the non-rotating cases (although, as underlined above,  already in models with no rotation, due to the importance of the convective core, stars are never very far from chemical homogeneity). As was the case in the non-rotating tracks, the O-type star phase corresponds to an upward displacement when time goes on in the HR diagram for the 150 $M_\odot$ model, while, it corresponds to a downwards displacement for the three more massive models. One notes finally that lower luminosities are reached by the rotating models at the end of their evolution (decrease by about 0.3 dex in luminosity, thus by a factor 2). This comes mainly because the rotating models enter earlier into their WR phase and thus lose more mass.

\begin{figure}
\center
\includegraphics[width=0.9\textwidth,clip=]{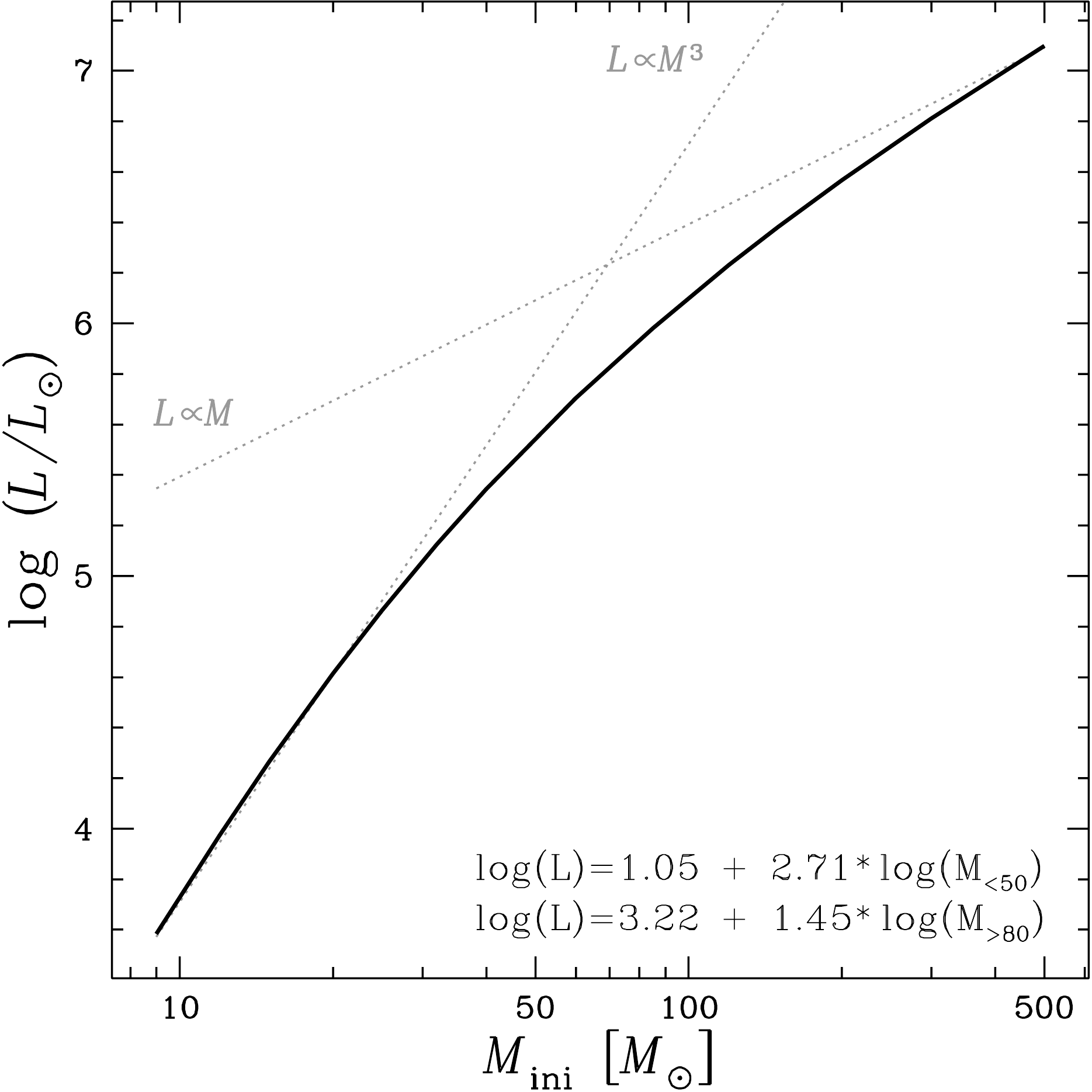}
\caption{ Mass-luminosity relation on the ZAMS for rotating models at solar metallicity. The formulae in the bottom right corner are linear fits for the mass ranges: 9--50\,$M_{\odot}$ and 80--500\,$M_{\odot}$. The non-rotating models have very similar properties on the ZAMS.
}\label{fig:LvsM}
\end{figure}

How does a change in metallicity alter the picture? 
When the metallicity decreases to Z=0.006 (see Fig.~\ref{fig:hrdlmcsmc}, {\it left}), as expected, tracks are shifted to higher luminosities and effective temperatures. 
In this metallicity range, all models evolve upwards during their O-type star phase in the HR diagram. This is an effect of the lower mass loss rates.

As was already the case at Z=0.014, rotation makes the star evolve nearly vertically in the HR diagram. One notes in this metallicity range, much more important effects of rotation than at Z=0.014, which is also expected, since at these lower metallicity, mass loss rates are smaller and rotational mixing more efficient. We note that most of the decrease in luminosity in the 500 $M_\odot$ solar mass model occurs during the WC phase in the Z=0.006 non-rotating model, while it occurs during the WNL phase in the rotating one. This illustrates the fact that rotational mixing, by creating a much larger H-rich region in the star, tends to considerably increase the duration of the WNL phase. One notes also that while the 150 $M_\odot$ model enters the WR phase only after the MS phase, the rotating model becomes a WR star before the end of the MS phase.

At the metallicity of the SMC (see Fig.~\ref{fig:hrdlmcsmc}, {\it right}), except for the 500 $M_\odot$, the tracks evolve horizontally after the end of the core H-burning phase (triangle in Fig.~\ref{fig:hrdlmcsmc}, {\it right}). The much lower mass loss rates are responsible for this effect.

\subsection{Lifetimes and mass-luminosity relation}

\begin{table*}
\caption{Properties of the hydrogen burning phase: initial properties of stellar models (columns 1-3), lifetime of H-burning and O-type star phase (4-5), average MS surface velocity (6) and properties at the end of the core H-burning phase (7-15). Masses are in solar masses, velocities are in km\,s$^{-1}$,
lifetimes are in $10^6$ years and abundances are surface abundances in mass fractions. The luminosity, $L$, is in log$_{10}(L/L_\odot )$ unit and the effective temperature, $T_\mathrm{eff}$, is in log$_{10}$ [K]. Note that the effective temperature given here includes a correction for WR stars to take into account the fact that their winds are optically thick as in \citet{ROTXI}.}\label{Table:endH}
\begin{tabular}{lccccccccccccccr}
\hline
 $M_{\rm ini}$  &$Z_{\rm ini}$  &$\frac{v_{\rm ini}}{v_\mathrm{crit}}$    &$\tau_H$   &$\tau_o$  &$\left<v_{\rm MS}\right>$
 &$M_\mathrm{H.b.}^\mathrm{end}$  &$^1$H  &$^4$He &$^{12}$C   &$^{14}$N &$^{16}$O &$T_\mathrm{eff}$ &$L$ &$\Gamma_\mathrm{Edd}$ \\
\hline
 120     &0.014   &0.0     &2.671   &2.592     &0.0   &63.7     &2.04e-1  &7.82e-1  &8.58e-5  &8.15e-3  &1.06e-4  &4.405 &6.334 &0.627\\
 150     &0.014   &0.0     &2.497   &2.348     &0.0   &76.3     &1.35e-1  &8.51e-1  &9.26e-5  &8.15e-3  &9.91e-5  &4.413 &6.455 &0.657 \\
 200     &0.014   &0.0     &2.323   &2.095     &0.0   &95.2     &7.51e-2  &9.11e-1  &9.93e-5  &8.14e-3  &9.23e-5  &4.405 &6.597 &0.687 \\
 300     &0.014   &0.0     &2.154   &1.657     &0.0   &65.2     &1.24e-3  &9.85e-1  &1.31e-4  &8.11e-3  &7.93e-5  &4.267 &6.401 &0.595 \\
 500     &0.014   &0.0     &1.990   &1.421     &0.0   &56.3     &2.20e-3  &9.84e-1  &1.26e-4  &8.12e-3  &8.03e-5  &4.301 &6.318 &0.568\\
 \hline
 120     &0.014   &0.4     &3.137   &2.270     &116.71 &34.6    &1.56e-3  &9.85e-1  &1.33e-4  &8.10e-3  &8.48e-5  &4.400  &6.018 &0.463 \\
 150     &0.014   &0.4     &2.909   &2.074     &101.24 &37.1    &1.80e-3  &9.85e-1  &1.30e-4  &8.11e-3  &8.41e-5  &4.387  &6.062 &0.479 \\
 200     &0.014   &0.4     &2.649   &1.830     &89.33  &40.0    &1.41e-3  &9.85e-1  &1.33e-4  &8.10e-3  &8.30e-5  &4.372  &6.110 &0.495 \\
 300     &0.014   &0.4     &2.376   &1.561     &61.16  &43.2    &1.85e-3  &9.85e-1  &1.33e-4  &8.10e-3  &8.23e-5  &4.356  &6.157 &0.511  \\
 500     &0.014   &0.4     &2.132   &1.377     &24.55  &48.1    &1.24e-3  &9.85e-1  &1.38e-4  &8.10e-3  &8.08e-5  &4.332  &6.221 &0.531 \\
 \hline
 120    &0.006    &0.0    &2.675   &2.682         &0.0  &79.0   &4.03e-1  &5.91e-1  &3.29e-5  &3.50e-3  &4.47e-5  &4.441 &6.391 &0.672 \\
 150    &0.006    &0.0    &2.492   &2.499         &0.0  &96.1   &3.28e-1  &6.67e-1  &3.58e-5  &3.50e-3  &4.25e-5  &4.483 &6.524 &0.709 \\
 500    &0.006    &0.0    &1.904   &1.636         &0.0  &238.8  &2.56e-2  &9.69e-1  &5.12e-5  &3.48e-3  &3.18e-5  &4.032 &7.094 &0.819 \\
\hline
 120     &0.006   &0.4   &3.140   &2.479         &208.55  &64.0 &1.70e-3  &9.92e-1  &6.06e-5  &3.47e-3  &3.04e-5  &4.387 &6.395 &0.597  \\
 150     &0.006   &0.4   &2.857   &2.172         &198.19  &71.3 &9.76e-4  &9.93e-1  &6.33e-5  &3.47e-3  &2.97e-5  &4.365 &6.455 &0.615  \\
 200     &0.006   &0.4   &2.590   &1.894         &193.05  &80.7 &1.22e-3  &9.93e-1  &6.29e-5  &3.47e-3  &2.95e-5  &4.339 &6.525 &0.638 \\
 300     &0.006   &0.4   &2.318   &1.619         &173.47  &85.8 &1.32e-3  &9.93e-1  &6.30e-5  &3.47e-3  &2.93e-5  &4.327 &6.559 &0.649 \\
 500     &0.006   &0.4   &2.077   &1.419         &116.76 &101.7 &1.37e-3  &9.93e-1  &6.37e-5  &3.47e-3  &2.89e-5  &4.291 &6.650 &0.676 \\
\hline
 150     &0.002   &0.4   &2.921   &2.567      &318.92  &128.8   &1.67e-3  &9.96e-1  &2.13e-5  &1.16e-3  &8.09e-6  &4.394 &6.780  &0.720 \\
 200     &0.002   &0.4   &2.612   &2.168      &333.43  &152.2   &1.31e-3  &9.97e-1  &2.26e-5  &1.16e-3  &7.82e-6  &4.363 &6.867  &0.743 \\
 300     &0.002   &0.4   &2.315   &1.801      &347.32  &176.2   &1.10e-3  &9.97e-1  &2.32e-5  &1.16e-3  &7.68e-6  &4.279 &7.067  &0.763 \\
 \hline
\end{tabular}
\end{table*}

\begin{table*}
\caption{Properties of the helium burning phase:
initial properties of stellar models (columns 1-3), age of star at the end of He-burning (4), average He-b. surface velocity (5) and properties at the end of the core He-burning phase (6-15).{ Abundances are given for the surface, except for $^{12}$C$_c$, which represents the central C abundance. Same units as in Table \ref{Table:endH}. ${\cal L}_{\rm CO}$ [$10^{50}\,\frac{{\rm g\,cm}^2}{\rm s}$] is the angular momentum contained in the CO core (Note that at this stage the angular velocity is constant in the CO core due to convective mixing).}}\label{Table:endHe}
\centering
\begin{tabular}{ccccccccccccccc}
\hline
 $M_{\rm ini}$  &$Z_{\rm ini}$  &$\frac{v_{\rm ini}}{v_\mathrm{crit}}$  &age$_\mathrm{He-b.}^\mathrm{end}$ &$\left<v_{\rm
 He-b.}\right>$ &$M_\mathrm{He-b.}^\mathrm{end}$   &$^4$He &$^{12}$C &$^{12}$C$_c$ &$^{16}$O &$^{22}$Ne &$T_\mathrm{eff}$ &$L$ &$\Gamma_\mathrm{Edd}$ & ${\cal L}_{\rm CO}$\\
\hline
 120  &0.014 &0.0 &3.003  &0.00    &30.9   &0.242  &0.458  &0.150  &0.281  &1.081e-02  &4.819  &6.117  &0.650 & 0    \\
 150  &0.014 &0.0 &2.809  &0.00    &41.3   &0.234  &0.436  &0.126  &0.312  &1.003e-02  &4.822  &6.278  &0.706 & 0    \\
 200  &0.014 &0.0 &2.622  &0.00    &49.4   &0.207  &0.408  &0.112  &0.366  &8.811e-03  &4.807  &6.377  &0.737 & 0    \\
 300  &0.014 &0.0 &2.469  &0.00    &38.2   &0.234  &0.443  &0.133  &0.305  &1.029e-02  &4.825  &6.236  &0.691 & 0    \\
 500  &0.014 &0.0 &2.314  &0.00    &29.8   &0.261  &0.464  &0.152  &0.257  &1.110e-02  &4.811  &6.095  &0.640 & 0    \\
 \hline
 120  &0.014 &0.4 &3.513  &1.58    &18.8   &0.292  &0.492  &0.195  &0.198  &1.196e-02  &4.806  &5.814  &0.533 & 1.91 \\
 150  &0.014 &0.4 &3.291  &1.18    &20.3   &0.286  &0.488  &0.187  &0.208  &1.184e-02  &4.808  &5.863  &0.551 & 1.91 \\
 200  &0.014 &0.4 &3.020  &0.50    &22.0   &0.277  &0.484  &0.180  &0.221  &1.172e-02  &4.812  &5.912  &0.570 & 1.37 \\
 300  &0.014 &0.4 &2.733  &0.13    &24.0   &0.270  &0.479  &0.172  &0.233  &1.151e-02  &4.814  &5.965  &0.589 & 0.75 \\
 500  &0.014 &0.4 &2.502  &0.03    &25.9   &0.269  &0.473  &0.164  &0.239  &1.140e-02  &4.811  &6.010  &0.606 & 0.28 \\
 \hline
 120  &0.006 &0.0 &2.993  &0.00    &54.2   &0.229  &0.391  &0.098  &0.372  &3.701e-03  &4.860  &6.424  &0.753 & 0    \\
 150  &0.006 &0.0 &2.845  &0.00    &59.7   &0.241  &0.370  &0.086  &0.380  &3.597e-03  &4.844  &6.474  &0.767 & 0    \\
 500  &0.006 &0.0 &2.182  &0.00    &94.7   &0.251  &0.392  &0.078  &0.349  &3.318e-03  &4.833  &6.711  &0.834 & 0    \\
 \hline
 120  &0.006 &0.4 &3.472  &6.84    &39.3   &0.294  &0.457  &0.132  &0.241  &4.709e-03  &4.387  &6.395  &0.692 & 16.2 \\
 150  &0.006 &0.4 &3.164  &3.67    &45.7   &0.310  &0.451  &0.122  &0.231  &4.701e-03  &4.824  &6.329  &0.767 & 14.7 \\
 200  &0.006 &0.4 &2.904  &1.33    &51.1   &0.303  &0.444  &0.114  &0.245  &4.547e-03  &4.825  &6.390  &0.738 & 9.98 \\
 300  &0.006 &0.4 &2.625  &0.35    &54.1   &0.291  &0.439  &0.110  &0.262  &4.433e-03  &4.830  &6.421  &0.748 & 5.18 \\
 500  &0.006 &0.4 &2.387  &0.13    &74.9   &0.330  &0.425  &0.090  &0.237  &4.356e-03  &4.790  &6.590  &0.798 & 4.83 \\
 \hline
 150  &0.002 &0.4 &3.193  &64.94   &106.7  &0.809  &0.153  &0.074  &0.035  &1.730e-03  &4.743  &6.766  &0.841 & 412.5\\
 200  &0.002 &0.4 &2.889  &29.88   &129.3  &0.880  &0.109  &0.066  &0.009  &1.777e-03  &4.789  &6.861  &0.863 & 355.6\\
 300  &0.002 &0.4 &2.585  &5.10    &149.8  &0.938  &0.058  &0.060  &0.001  &1.798e-03  &4.833  &6.933  &0.880 & 156.8\\
\hline
\end{tabular}
\end{table*}

In Tables \ref{Table:endH} and \ref{Table:endHe}, we 
provide ages at the end of core hydrogen burning and core helium 
burning, respectively. We see that the MS lifetime of non-rotating models at solar metallicity ranges from 2.67\,Myr to 1.99\,Myr for initial masses ranging from 120 to 500 $M_\odot$
showing the well known fact that VMS have a very weak lifetime dependence on their initial mass.

The mass-luminosity relation on the ZAMS for rotating massive stars at solar 
composition is shown in Fig.~\ref{fig:LvsM}. The relation ($L \propto M^{\alpha}$) is steep for low and intermediate-mass stars ($\alpha \sim$ 3 for $10 < M/M_{\odot} < 20$) and flattens for VMS ($\alpha \sim 1.3$ for $200 < M/M_{\odot} < 500$).
This flattening is due to the increased radiation pressure relative to gas pressure in massive stars. Since the lifetime of a star is roughly $M/L$, we get that for VMS $\tau \propto M/L \propto M^{-0.3}$.

The H-burning (and total) lifetimes of VMS are lengthened by 
rotation as in lower mass stars. Differences in the H-burning lifetimes 
of rotating and non-rotating 150 $M_\odot$ models at solar metallicity 
are $\sim$ 14\%. The effects of metallicity on the lifetimes are generally very small. The small differences in total lifetimes are due to different mass loss at different metallicities.

\subsection{Mass loss by stellar winds}

\begin{figure}
\centering
\includegraphics[width=0.7\textwidth,clip=]{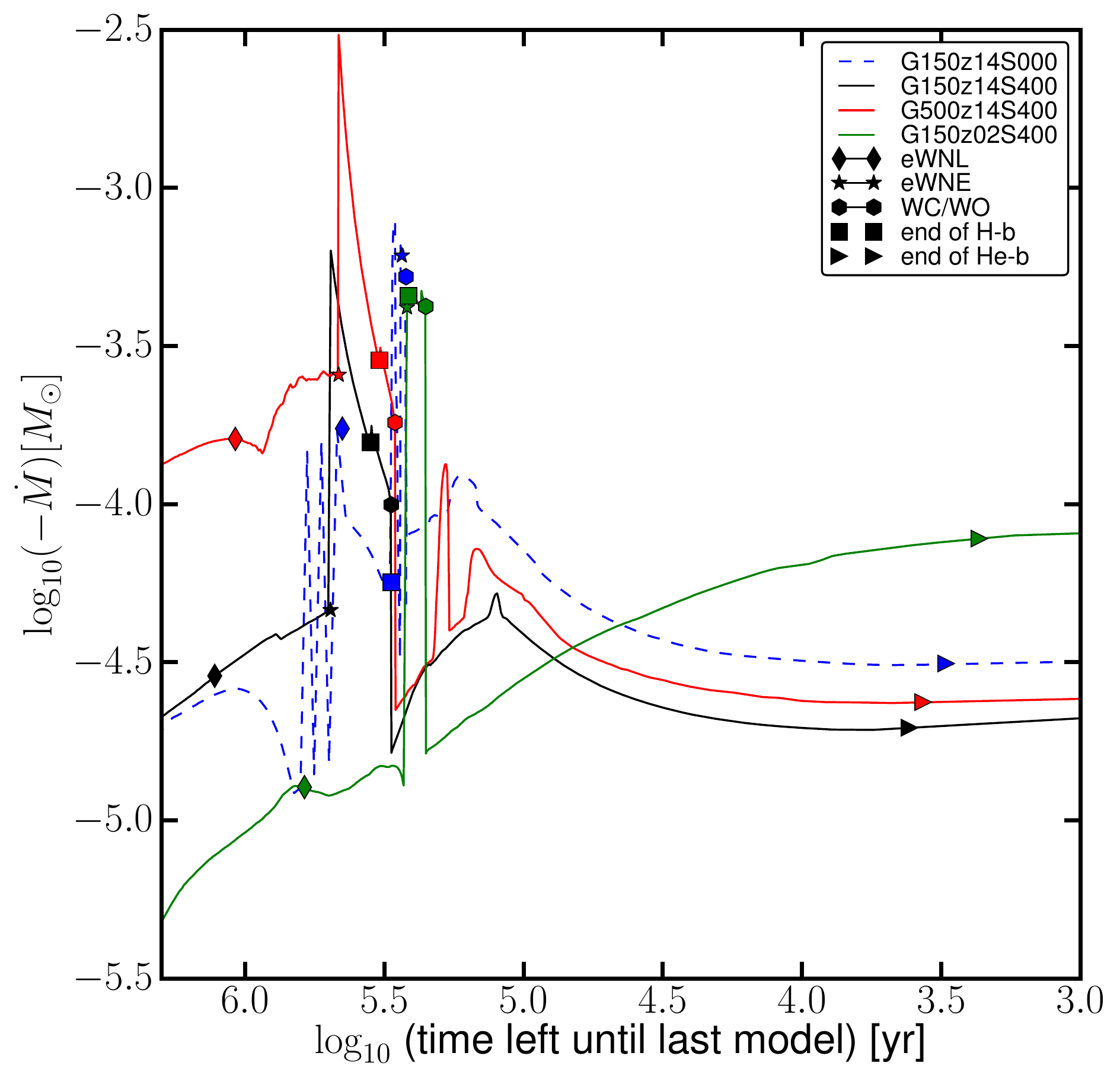}
\caption{Evolution of the mass loss rate as a function of time left until last model (log scale) for the rotating 500 $M_\odot$ model (solid-red), the rotating 150 $M_\odot$ model (solid-black), the non-rotating (dashed) 150 $M_\odot$ model at solar metallicity, and the rotating 150 $M_\odot$ model at SMC metallicity (solid-green). The diamonds indicate the start of the eWNL phase, the stars the start of the eWNE phase and hexagons the start of the WC/WO phase. The squares and triangles indicate the end of H-b. and He-b. phases, respectively.}
\label{fig:massloss_yr}
\end{figure}

\begin{table*}
\caption{Mass loss properties: Total mass of the models at various stages (columns 1-5), and average mass loss rates $\left<\dot{M}\right>$ during the O-type and eWNE phases (6,7). Masses are in solar mass units and mass loss rates are given in $M_\odot$yr$^{-1}$. }\label{mdotrates}
\centering
 \begin{tabular}{ccccc|cc}
\hline
ZAMS &start eWNL & start eWNE & start WC & final &$\left<\dot{M}_\mathrm{Vink}\right>$ &$\left<\dot{M}_\mathrm{eWNE}\right>$ \\
\hline
&\multicolumn{3}{c}{{Z=0.014, $v/v_{crit}=0.0$}} &&&  \\
120   &69.43  &52.59  &47.62 &30.81 &2.477e-05 &3.638e-04  \\
150   &88.86  &66.87  &61.20 &41.16 &3.274e-05 &6.107e-04  \\
200   &121.06 &91.20  &83.85 &49.32 &4.618e-05 &1.150e-03  \\
300   &184.27 &130.47 &52.05 &38.15 &8.047e-05 &8.912e-04  \\
500   &298.79 &169.50 &45.14 &29.75 &1.736e-04 &9.590e-04  \\
\hline
&\multicolumn{3}{c}{{Z=0.014, $v/v_{crit}=0.4$}} &&&  \\
120   &88.28 &69.54   &27.43 &18.68 &1.675e-05 &2.057e-04   \\
150   &106.64 &80.88  &29.49 &20.22 &2.467e-05 &2.640e-04  \\
200   &137.52 &98.75  &31.84 &21.93 &3.985e-05 &3.564e-04  \\
300   &196.64 &129.10 &34.45 &23.93 &7.559e-05 &5.160e-04  \\
500   &298.42 &174.05 &38.30 &25.83 &1.594e-04 &7.901e-04  \\
\hline
&\multicolumn{3}{c}{{Z=0.006, $v/v_{crit}=0.0$}} &&&  \\
120   &74.30  &57.91  &56.91  &54.11  &2.140e-05 &3.272e-04  \\
150   &94.18  &74.20  &71.75  &59.59  &2.839e-05 &5.038e-04  \\
500   &332.68 &250.64 &197.41 &94.56  &1.304e-04 &3.334e-03   \\
\hline
&\multicolumn{3}{c}{{Z=0.006, $v/v_{crit}=0.4$}} &&&  \\
120   &100.57 &90.78  &54.43 &39.25 &9.429e-06 &3.219e-04 \\
150   &125.79 &111.84 &60.75 &45.58 &1.367e-05 &4.418e-04 \\
200   &166.81 &144.86 &66.25 &51.02 &2.180e-05 &6.257e-04 \\
300   &247.07 &207.10 &73.11 &54.04 &4.166e-05 &9.524e-04 \\
500   &397.34 &315.51 &86.10 &74.75 &9.194e-05 &1.685e-03  \\
\hline
&\multicolumn{3}{c}{{Z=0.002, $v/v_{crit}=0.4$}} &&&  \\
150  &135.06 &130.46 &113.51 &106.50 &6.661e-06 &4.485e-04 \\
200  &181.42 &174.18 &137.90 &129.21 &9.902e-06 &6.631e-04 \\
300  &273.18 &260.81 &156.14 &149.70 &1.730e-05 &1.040e-03 \\
\hline
 \end{tabular}
\end{table*}

Mass loss by stellar winds is a key factor governing the evolution of VMS. This comes from the very high luminosities reached by these objects. 
For example, the luminosity derived for R136a1 is about 10 million times that of our sun.

For such luminous objects,
winds will be very powerful at all evolutionary stages, so while
early main-sequence VMS are formally O-type stars from an evolutionary 
perspective, their spectral appearance may be closer to Of or Of/WN at 
early phases \citep{Crowther12}.

Table \ref{mdotrates} gives the total mass at the start and end of the evolution \footnote{the models have been evolved beyond the end of core He-burning and usually until oxygen burning, thus very close to the end of their life \citep[see][for full details]{Liza13}} as well as at the transitions between the different WR phases in columns 1 to 5. 
The average mass loss rates during the O-type and eWNE phases (the phase during which the mass loss rates are highest)
are given in columns 6 and 7, respectively. 

The evolution of the mass loss rates for various models are shown in Fig.~\ref{fig:massloss_yr}.
Following the evolution from left to right for the 150 $M_\odot$ model at solar metallicity (solid-black), mass loss rates slowly increase at the start of the O-type phase with mass loss rates between $10^{-5}\,M_\odot$\,yr$^{-1}$ (absolute values for the mass loss rates, $-\dot{M}$, are quoted in this paragraph) and $10^{-4.5}\,M_\odot$\,yr$^{-1}$. 
If a bi-stability limit is encountered during the MS phase, as is the case in the non-rotating 150 $M_{\odot}$ model, mass loss rates can vary significantly over a short period of time and mass loss peaks reach values higher than $10^{-4}\,M_\odot$\,yr$^{-1}$. The highest mass loss rate is encountered at the start of the eWNE phase (star symbols) with values in excess of $10^{-3}\,M_\odot$\,yr$^{-1}$ (note that the mass loss rate in the non-rotating model has a peak at the end of the H-burning phase. phase due to the star reaching temporarily cooler effective temperatures). Such high mass loss rates quickly reduce the mass and luminosity of the star and thus the mass loss rate also decreases quickly during the eWNE phase. During the WC/WO phase, mass loss rates are of the same order of magnitude as during the O-type phase.

Comparing the rotating 500 and 150 $M_\odot$ model at solar metallicity (solid black and red), we see that more massive stars start with higher mass loss rates but converge later on to similar mass loss rates since the total mass of the models converges to similar values (see Table \ref{mdotrates}). 

Comparing the SMC and solar metallicity 150 $M_\odot$ rotating models, we can clearly see the metallicity effect during the O-type star phase. During the eWNE phase, mass loss rates are similar and in the WC/WO, mass loss rates in the SMC model are actually higher since the total mass in that model remained high in contrast with solar metallicity models.

Table \ref{mdotrates} also shows the relative importance of the mass lost during the various phases and how their importance changes as a function of metallicity. Even though mass loss is the strongest during the eWNE phase, significant amount of mass is lost in all phases.

\subsection{Mass loss rates and proximity of the Eddington limit}

\begin{figure}
\centering
\includegraphics[width=0.8\textwidth,clip=]{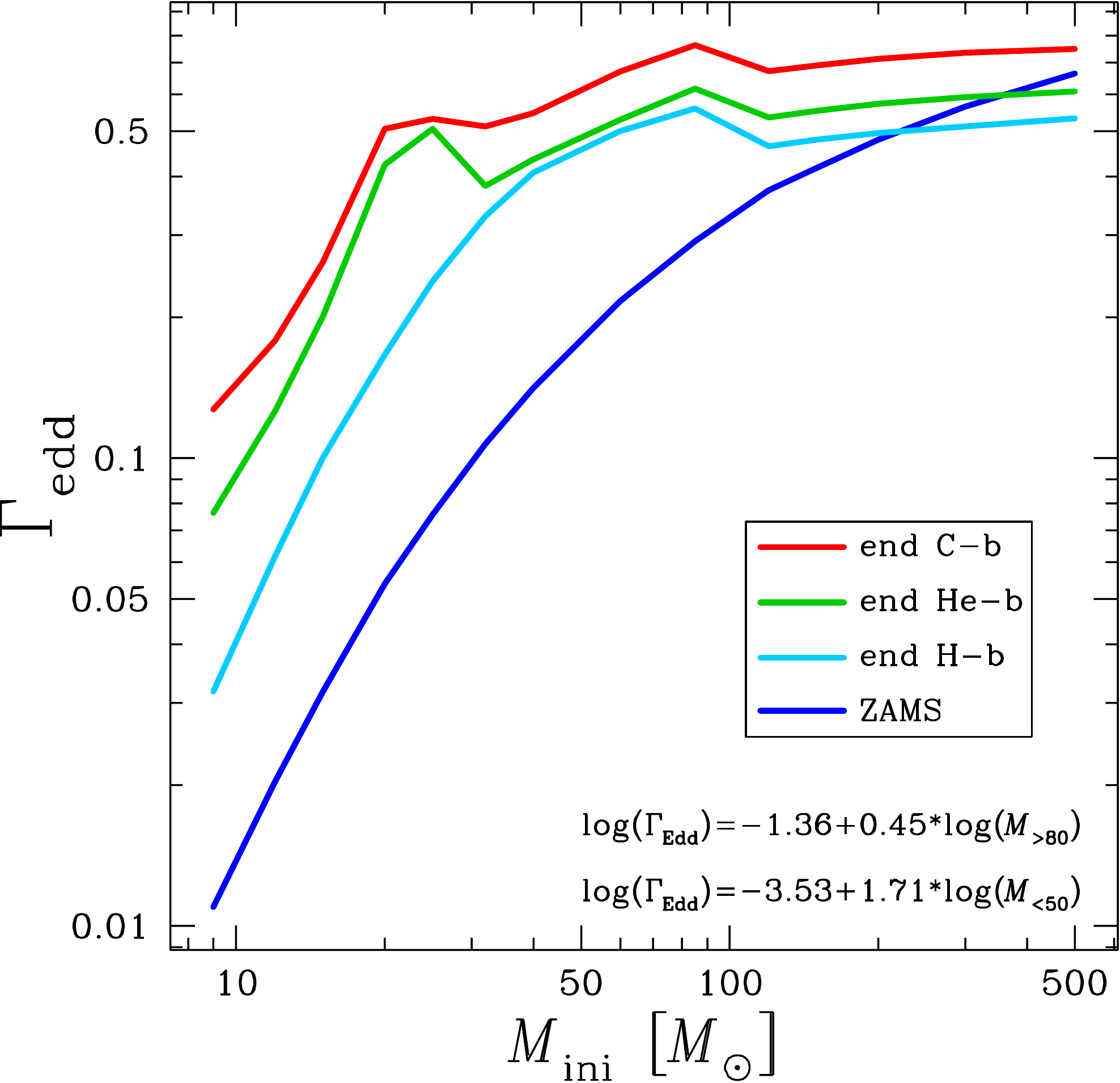}
\caption{Eddington parameter, $\Gamma_{\rm edd}$ for rotating models at solar metallicity. $\Gamma_{\rm edd}$ is plotted on the ZAMS (blue line) and the end of H-(light blue), He-(green) and C-burning (red) phases. Except for the 300 and 500 $M_\odot$ models, $\Gamma_{\rm edd}$ increases throughout the evolution. At solar metallicity, the highest value (close to 0.8) is actually reached by the 85 $M_\odot$ model at the end of its evolution. This could lead to significant mass loss shortly before the final explosion in a model that ends as a WR star and potentially explain supernova surrounded by a thick circumstellar material without the need for the star to be in the luminous variable phase.
The formulae in the bottom right corner are linear fits for the mass ranges: 9--50\,$M_{\odot}$ and 80--500\,$M_{\odot}$. The non-rotating models have very similar properties on the ZAMS.
}\label{fig:edd_mass}
\end{figure}

\citet{VMA11} 
suggest enhanced
mass-loss rates \citep[with respect to][used in the models presented here]{VN01} for stars with high 
Eddington parameters \citep[$\Gamma_{\rm e} \geq$ 0.7), see Eqn.\,1 in][for the exact definition of $\Gamma_\text{e}$]{VMA11} that they attribute to the 
Wolf-Rayet stage. In order to know whether higher mass loss rates near the Eddington limit could have an impact on the present result, we discuss here the proximity of our models to the Eddington limit.

Figure \ref{fig:edd_mass} shows the Eddington parameter, $\Gamma_\text{Edd}=L/L_\text{Edd}=\kappa L / (4\pi cGM)$, as a function of the initial mass of our models at key stages.
Since the Eddington parameter, $\Gamma_\text{Edd}$ scales with $L/M$, the curve for $\Gamma_\text{Edd}$ also flattens for VMS.
The ZAMS values for $\Gamma_\text{Edd}$ range between $0.4 - 0.6$, so well below the Eddington limit, $\Gamma_\text{Edd} = 1$, and below the limiting value of $\Gamma_\text{e}=0.7$ where enhanced mass-loss rates are expected according to \citet{VMA11}. 

\begin{figure}
\centering
\includegraphics[width=0.8\textwidth,clip=]{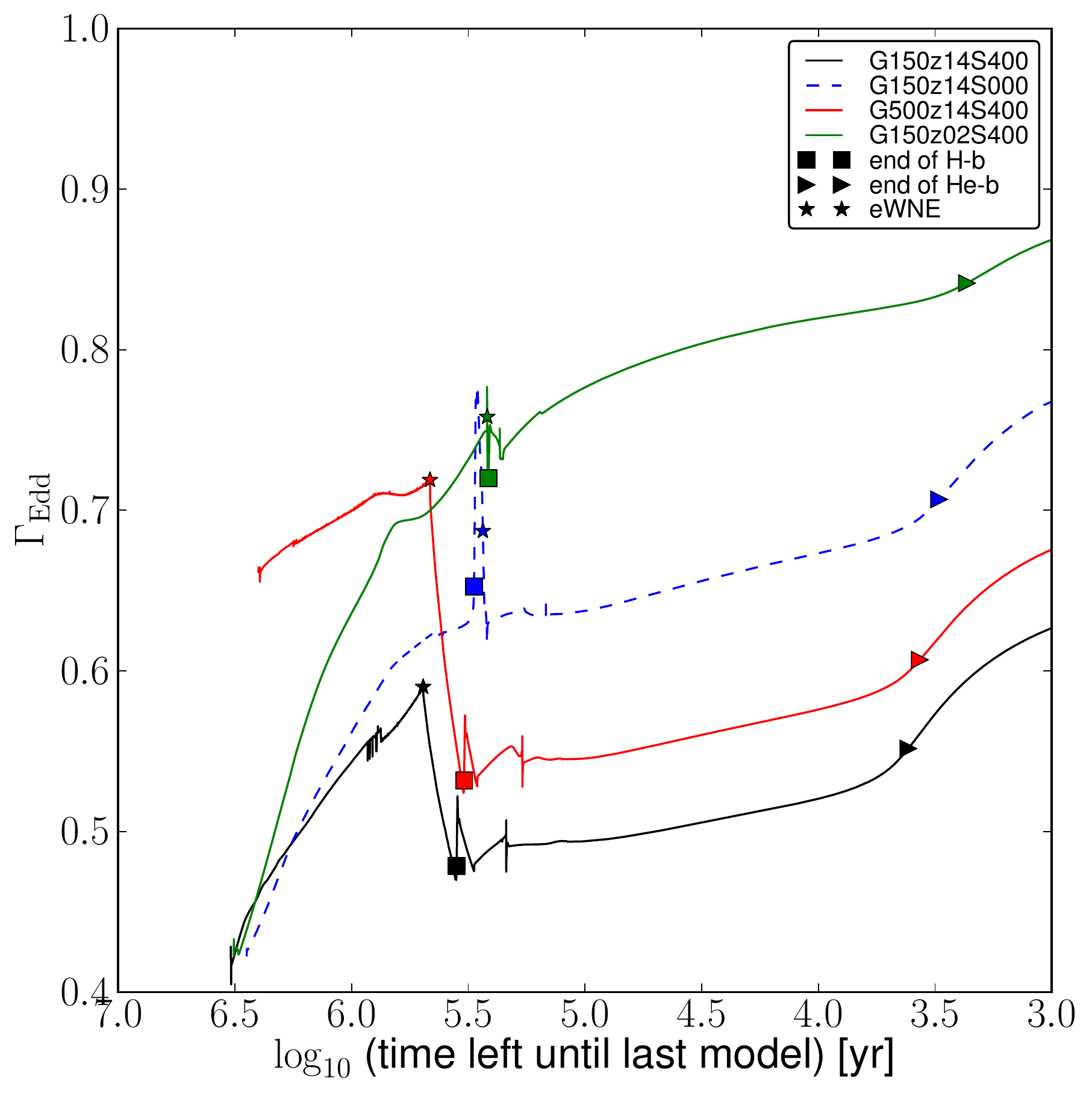}
\caption{Evolution of the Eddington parameter, $\Gamma_\text{Edd}$, as a function of time left until last model (log scale) for the rotating 500 $M_\odot$ model (solid-red), the rotating 150 $M_\odot$ model (solid-black), the non-rotating (dashed) 150 $M_\odot$ model at solar metallicity, and the rotating 150 $M_\odot$ model at SMC metallicity (solid-green). The stars indicate the start of the eWNE phase. The squares and triangles indicate the end of H-b. and He-b. phases, respectively.
}\label{fig:edd_yr}
\end{figure}

How does $\Gamma_\text{Edd}$ change during the lifetime of VMS? {Fig.~\ref{fig:edd_yr} presents the evolution of $\Gamma_\text{Edd}$ for a subset of representative models. The numerical values for each model are given at key stages in Tables \ref{Table:endH} and \ref{Table:endHe}. Since $\Gamma_\text{Edd}\propto \kappa L/M$, an increase in luminosity and a decrease in mass both lead to higher $\Gamma_\text{Edd}$. Changes in effective temperature and chemical composition affect the opacity and also lead to changes in $\Gamma_\text{Edd}$.}

{In rotating models at solar metallicity, $\Gamma_\text{Edd}$ slowly increases until the start of the eWNE phase. This is mainly due to the increase in luminosity and decrease in mass of the model. At the start of the eWNE phase, mass loss increases significantly. This leads to a strong decrease in the luminosity of the model and as a result $\Gamma_\text{Edd}$ decreases sharply.}

During the WC/WO phase, mass loss rates being of similar values as during the O-type star phase, $\Gamma_\text{Edd}$ increases again gradually.

We can see that, at solar metallicity, $\Gamma_\text{Edd}$ rarely increases beyond 0.7 even in the 500 $M_\odot$ model. 
There are nevertheless two interesting cases in which values above 0.7 are reached. The first case is during the advanced stages. At this stage, mass loss does not have much time to change the total mass of the star (it is mostly changes in effective temperature and to a minor extent in luminosity that influence the increase in $\Gamma_\text{Edd}$). This may nevertheless trigger
instabilities resulting in strong mass loss episodes.
This may have consequences for the type of SN
event that such a star will produce and may be a
reason why the explosion of VMS may look like as if they
had happened in an environment similar to those observed around Luminous Blue Variable. The second case is at low metallicity, as highlighted by the 150 $M_\odot$ model at SMC metallicity. Indeed, values above 0.7 are reached before the end of the MS (square symbol). Mass loss prescriptions such as the ones of \citet{VMA11} and \citet{MGME12} may thus play an important role on the fate of VMS. The non-rotating model has a different mass loss history (see Fig.~\ref{fig:massloss_yr}), which explains the slightly different evolution of $\Gamma_\text{Edd}$ near the end of the main sequence.

\subsection{Evolution of the surface velocity}\label{vsurf}

\begin{figure*}
\centering
\includegraphics[width=0.5\textwidth,clip=]{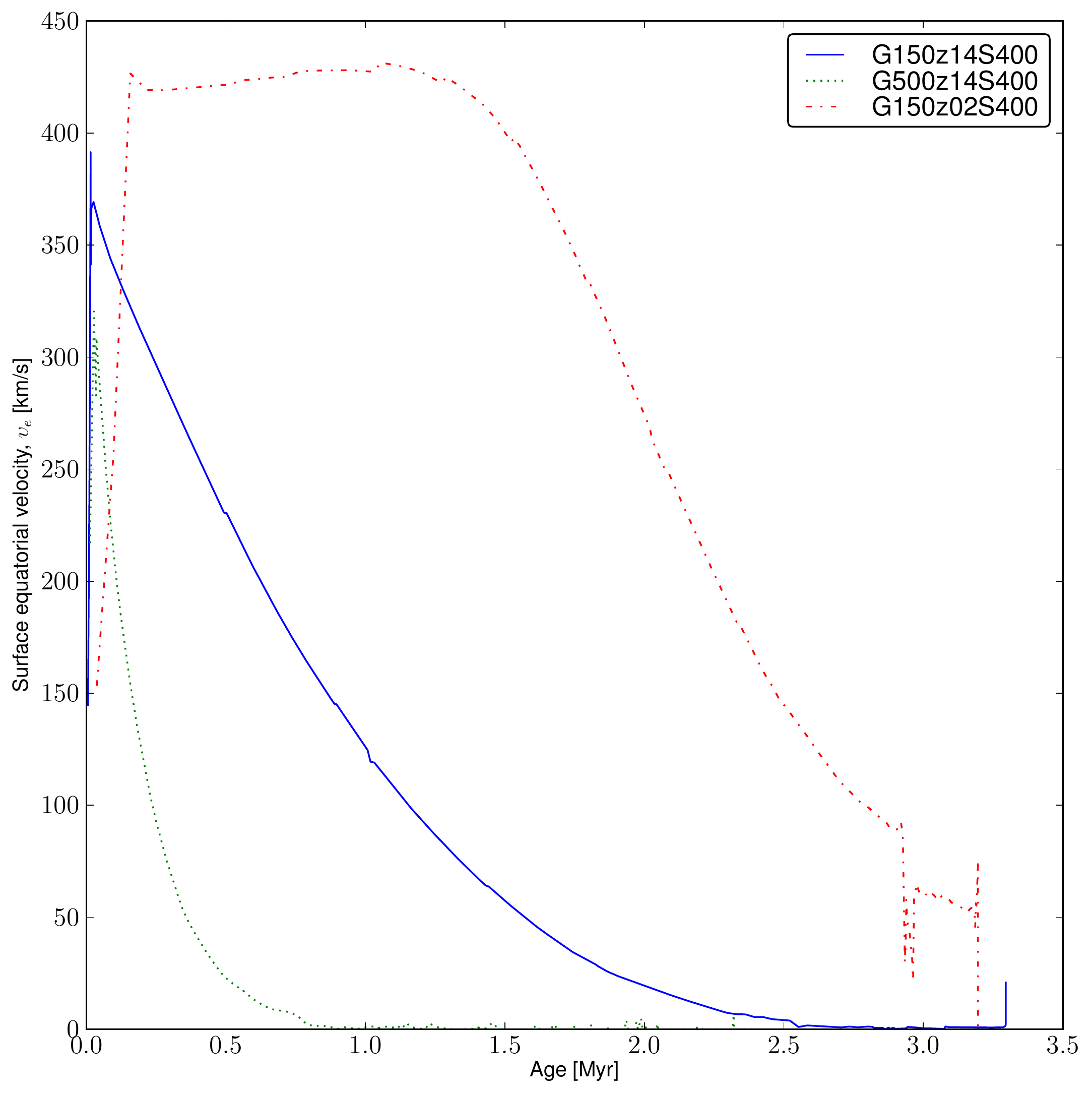}\includegraphics[width=0.5\textwidth,clip=]{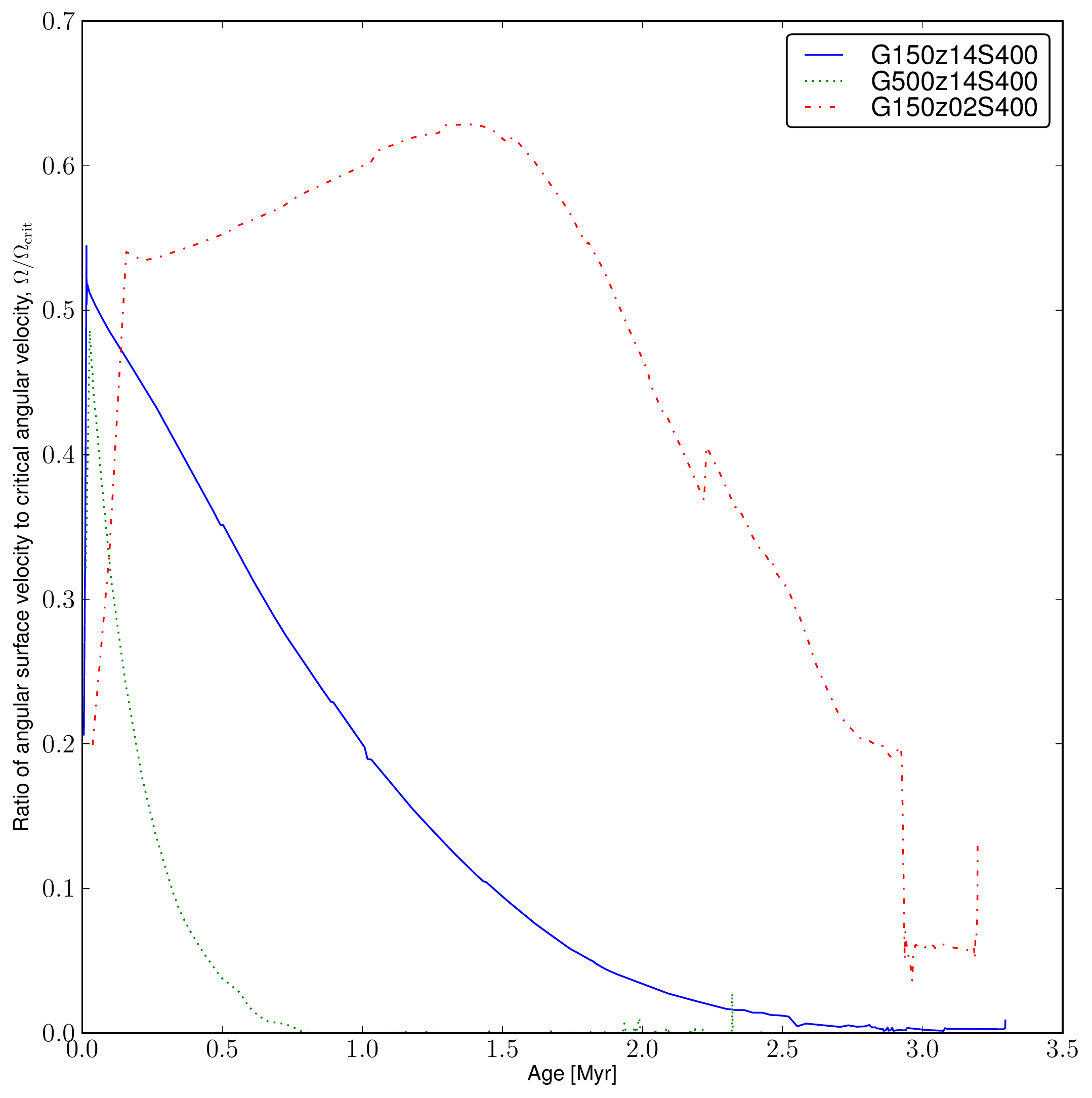} \\
\caption{Evolution of surface equatorial velocity ({\it left}) and ratio of the surface angular velocity to the critical angular velocity ({\it right}) for the rotating solar metallicity 150 and 500 $M_\odot$ and SMC 150 $M_\odot$ models as a function of age of the star.}\label{fig:vsurf}
\end{figure*}
The surface velocity of stars is affected by several processes. Contraction or expansion of the surface respectively increases and decreases the surface velocity due to the conservation of angular momentum. Mass loss removes angular momentum and thus decreases the surface velocity. Finally internal transport of angular momentum generally increases the surface velocity. As shown in Fig.\,\ref{fig:vsurf} (left panel), at solar metallicity, the surface velocity rapidly decreases during the main sequence due to the strong mass loss over the entire mass range of VMS. At SMC metallicity, mass loss is weaker than at solar metallicity and internal transport of angular momentum initially dominates over mass loss and the surface velocity increases during the first half of the MS phase. During this time, the ratio of surface velocity to critical velocity also increases up to values close to 0.7 \citep[note that the models presented include the effect of the luminosity of the star when determining the critical rotation as described in ][]{ROTVI}. However, as the evolution proceeds, the luminosity increases and 
mass loss eventually starts to dominate and the surface velocity and its ratio to critical rotation both decrease for the rest of the evolution. SMC stars thus never reach critical rotation. The situation at very low and zero metallicities has been studied by several groups \citep[see][and references therein]{H07,ES08,YDL12,CE12}. If mass loss becomes negligible, then the surface velocity reaches critical rotation for a large fraction of its lifetime, which probably leads to mechanical mass loss along the equator. The angular momentum content in the core of VMS stars is discussed further in Sect. \ref{grb}.

\section{WR stars from VMS}
\begin{figure}
\centering
\includegraphics[width=0.8\textwidth,clip=]{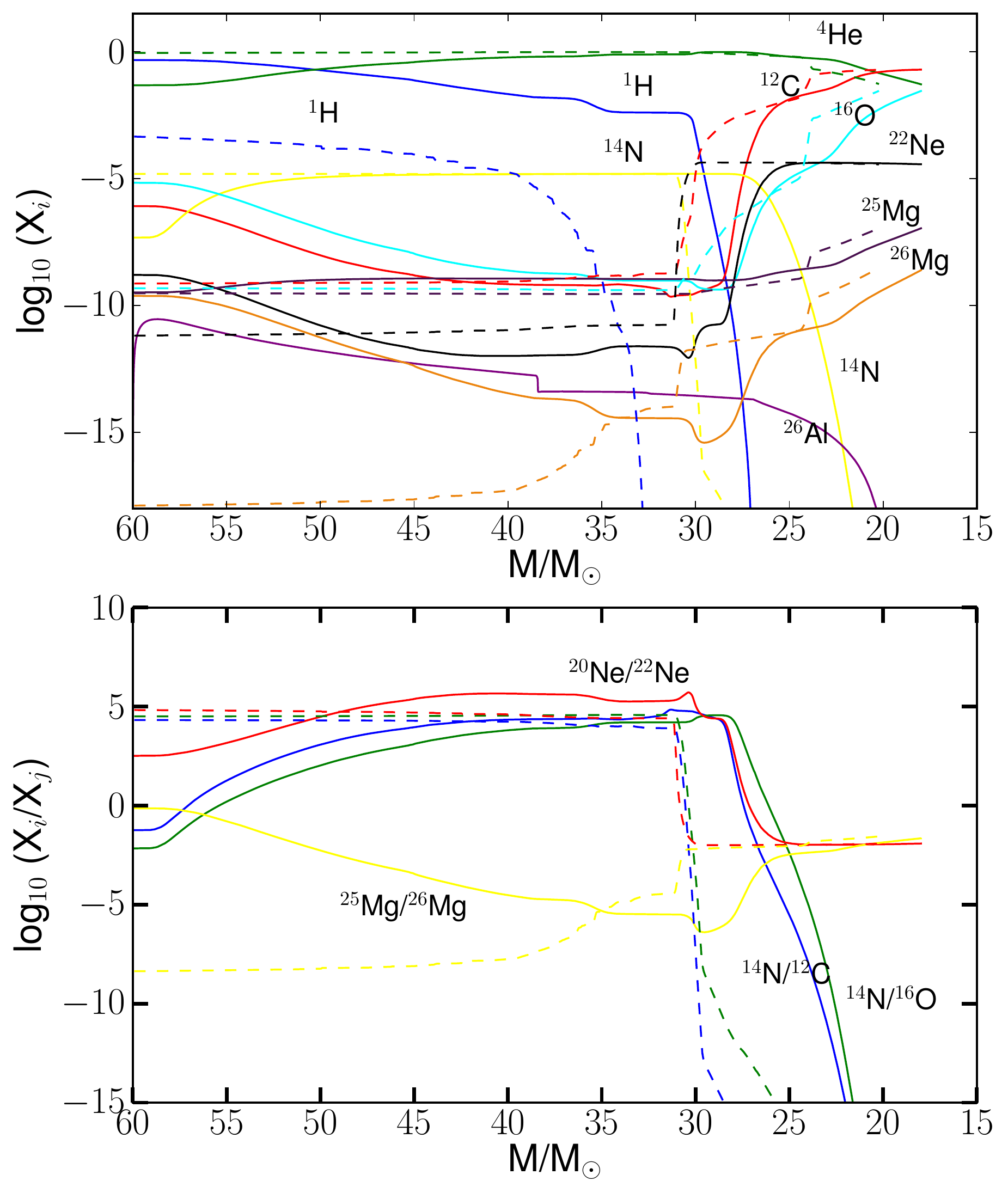} \\
\caption{
Evolution of surface abundances of the solar metallicity rotating 150 $M_\odot$ (solid) and 60 $M_\odot$ (dashed) rotating solar $Z$ models as a function of total mass (evolution goes from left to right since mass loss peels off the star and reduces the total mass). The top panel shows individual abundances while the bottom panel shows abundance ratios.
}\label{fig:surface_solar}
\end{figure}

Figure \ref{fig:surface_solar} presents the evolution of the surface abundances as a function of the total mass for the solar metallicity rotating models of 150 and 60 $M_\odot$.
This figure shows how the combined effects of mass loss and internal mixing change their surface composition.
Qualitatively there are no big differences between the 60 and 150 $M_\odot$ models.
Since the 150 $M_\odot$ has larger cores, the transition to the various WR stages occurs at larger total masses compared to the 60 $M_\odot$ model. It thus confirms the general idea that a more massive (thus more luminous) WR star originates from a more massive O-type star. Figure \ref{fig:surface_solar} shows that all abundances and abundance ratios are very similar for a given WR phase. it is therefore not easy to distinguish a WR originating from a VMS from its surface chemical composition (however see below).

\begin{table*}
\caption{Lifetimes of the various phases in units of years.}\label{wrlifetime}
 \begin{tabular}{ccccccccccccr}
\hline
$M_{\rm ini}$  &$Z_{\rm ini}$  &$\frac{v_{\rm ini}}{v_\mathrm{crit}}$        &O-star     &WR            &WNL        &WNE        &WN/WC      &WC (WO)            \\
\hline
120    &0.014	 &0      &2.151e06    &3.959e05    &1.150e05   &9.390e03   &2.675e02   &2.715e05          \\
150    &0.014	 &0      &2.041e06    &4.473e05    &1.777e05   &5.654e03   &7.120e02   &2.639e05          \\
200    &0.014	 &0      &1.968e06    &5.148e05    &2.503e05   &1.773e03   &4.576e02   &2.626e05       \\
300    &0.014	 &0      &1.671e06    &8.014e05    &5.051e05   &9.217e03   &2.735e03   &2.870e05          \\
500    &0.014	 &0      &1.286e06    &8.848e05    &5.804e05   &1.079e04   &3.279e03   &2.935e05       \\
\hline
120	&0.014   &0.4    &2.289e06    &1.227e06    &8.790e05   &4.118e04   &4.008e03   &3.076e05          \\
150	&0.014   &0.4    &2.105e06    &1.189e06    &8.567e05   &2.579e04   &3.649e03   &3.068e05          \\
200	&0.014   &0.4    &1.860e06    &1.164e06    &8.375e05   &2.242e04   &3.153e03   &3.042e05          \\
300	&0.014   &0.4    &1.585e06    &1.152e06    &8.315e05   &1.897e04   &2.897e03   &3.015e05          \\
500	&0.014   &0.4    &1.422e06    &1.083e06    &7.663e05   &1.830e04   &2.899e03   &2.990e05          \\
\hline
120     &0.006	 &0      &2.222e06    &2.964e05    &2.043e05   &1.302e02   &6.025e02   &9.202e04          \\
150     &0.006	 &0      &2.028e06    &3.320e05    &1.579e05   &1.211e03   &2.921e02   &1.728e05          \\
500     &0.006	 &0      &1.388e06    &5.362e05    &2.690e05   &5.211e03   &1.350e03   &2.620e05          \\
\hline
120    &0.006    &0.4    &2.513e06    &9.624e05    &6.776e05   &1.601e04   &3.386e03   &2.687e05          \\
150    &0.006    &0.4    &2.188e06    &9.789e05    &6.912e05   &2.172e04   &2.336e03   &2.660e05          \\
200    &0.006    &0.4    &1.922e06    &9.848e05    &7.073e05   &1.347e04   &2.757e03   &2.640e05          \\
300    &0.006    &0.4    &1.644e06    &9.838e05    &7.033e05   &1.600e04   &9.744e02   &2.644e05          \\
500    &0.006    &0.4    &1.461e06    &9.283e05    &6.647e05   &9.312e03   &6.853e02   &2.542e05          \\
\hline
150    &0.002    &0.4    &2.583e06    &6.119e05    &3.691e05   &8.459e03   &4.874e03   &2.343e05          \\
200    &0.002    &0.4    &2.196e06    &6.926e05    &4.524e05   &1.019e04   &2.709e03   &2.300e05          \\
300    &0.002    &0.4    &1.827e06    &7.602e05    &5.186e05   &1.317e04   &1.289e03   &2.283e05          \\
\hline
 \end{tabular}
\end{table*}

\begin{figure*}
\center
\includegraphics[width=0.5\textwidth,clip=]{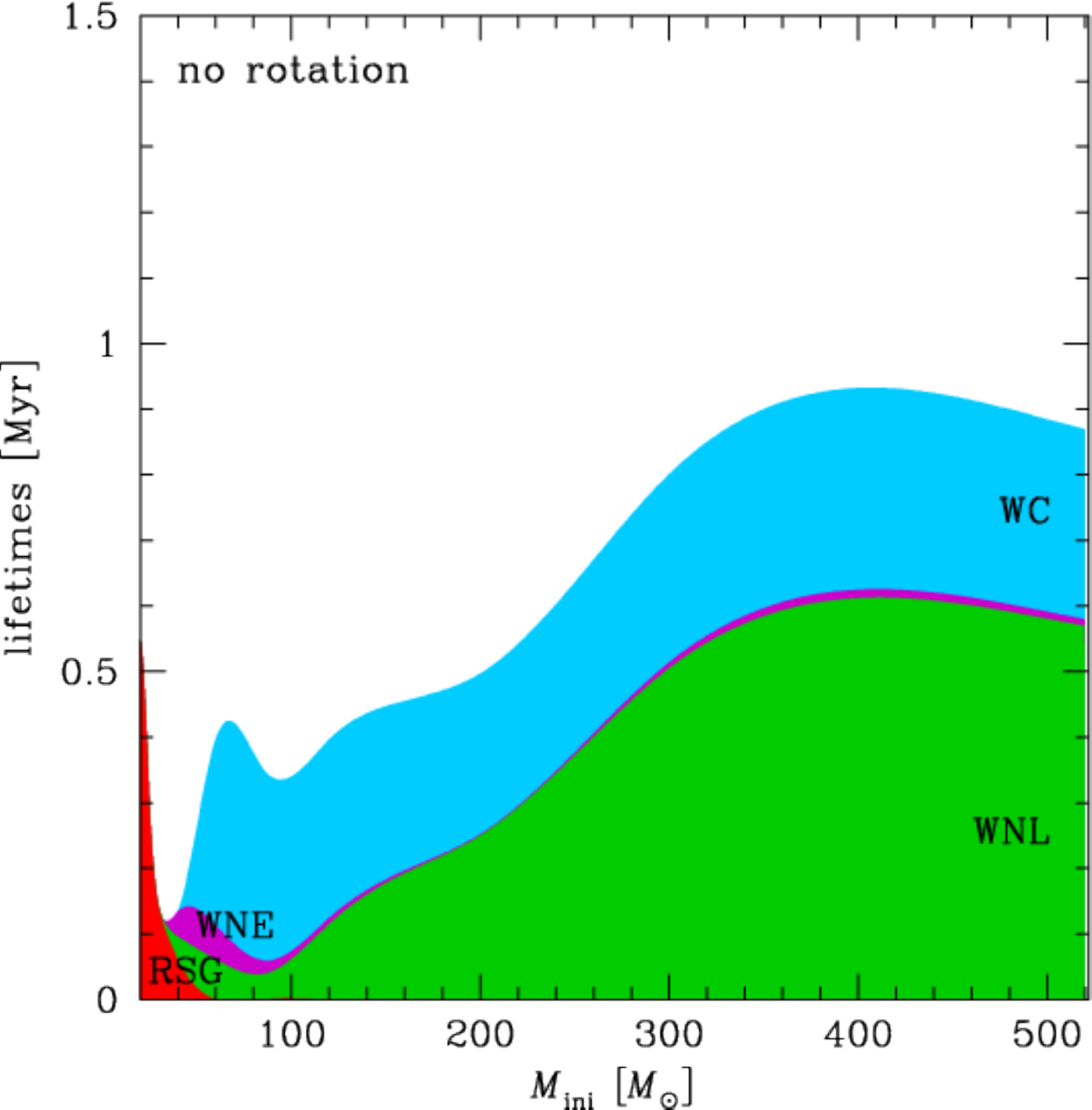}\includegraphics[width=0.5\textwidth,clip=]{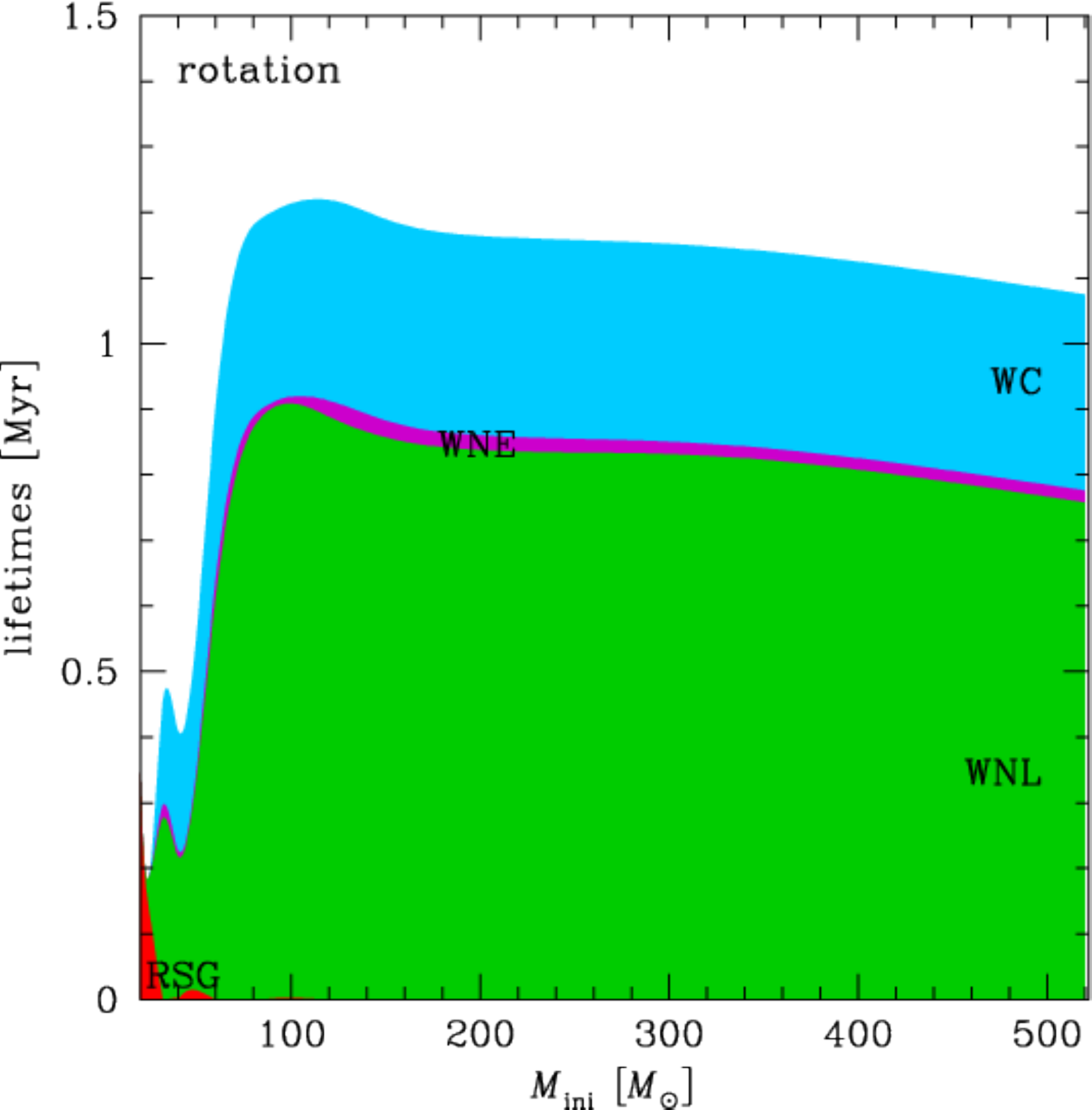}
\caption{Lifetimes of the RSG phase and of the different WR phases for the solar metallicity non-rotating ({\it left}) and rotating ({\it right}) models. Lifetimes are piled up. For example, the lifetime of the WNE phase extent corresponds to the height of the purple area.}
\label{fig:wrlife}
\end{figure*}

We present in Table \ref{wrlifetime} the lifetimes of the different WR phases
through which all our VMS models evolve.
At solar metallicity, the WR phase of non-rotating stellar models for masses between 150 and 500 $M_\odot$
covers between 16 and 38\% of the total stellar lifetime. This is a significantly larger proportion than for masses
between 20 and 120 $M_\odot$, where the WR phase covers only 0-13\% percents of the total stellar lifetimes.
At the LMC metallicity, the proportion of the total stellar lifetime spent as a WR phase for VMS decreases
to values between 12\% (150 $M_\odot$) and 25\% (500 $M_\odot$). 

Figure~\ref{fig:wrlife} shows how these lifetimes vary as a function of mass for
non-rotating and rotating solar metallicity models.
Looking first at the non-rotating models (Fig.~\ref{fig:wrlife}, {\it left}), we see that the very massive stars
(above 150 $M_\odot$) have WR lifetimes between 0.4 and nearly 1 My.
The longest WR phase is the WNL phase since these stars spend a large fraction of H-burning in this phase. The duration
of the WC phases of VMS is not so much different from those of stars in the mass range 
between 50 and 120 $M_\odot$. 

Rotation significantly increases the WR lifetimes. Typically, the WR phase of rotating stellar models for masses between 150 and 500 $M_\odot$
covers between 36 and 43\% of the total stellar lifetime.
The increase is more important
for the lower mass range plotted in the figures. This reflects the fact that
for lower initial mass stars, mass loss rates are weaker and thus the mixing
induced by rotation has a greater impact. We see that this increase is mostly due
to longer durations for the WNL phase, the WC phase duration remaining more or less
constant for the whole mass range between 50 and 500 $M_\odot$ as was the case
for the non rotating models. Rotation has qualitatively similar effects at the LMC metallicities.

Would the account of the VMS stars in the computation of the number ratios of WR to O-type stars and on the WN/WC
ratios have a significant effect? The inclusion of VMS is marginal at solar metallicity, since the durations are only affected
by a factor 2. Convoluted with the weighting of the initial mass function (IMF), WR stars originating from VMS only represent $\sim 10\%$ of the whole
population of WR stars \citep[using a][IMF]{S55} originating from single stars. However, the situation is different at SMC metallicity. Due to the weakness of the stellar winds, single stellar models below $120\, M_
\odot$ at this $Z$ do not produce any WC or WO stars (Georgy et al. in prep.). In that case, we expect that the few WC/WO
stars observed at low metallicity come from VMS, or from the binary channel \citep{EIT08}.
In starburst regions, the detection of WR stars at very young ages would also be an indication that they come from VMS, as
these stars enter the WR phase before their less massive counterparts, and well before WRs coming from the binary
channel.
\begin{figure}
\center
\includegraphics[width=0.6\textwidth,clip=]{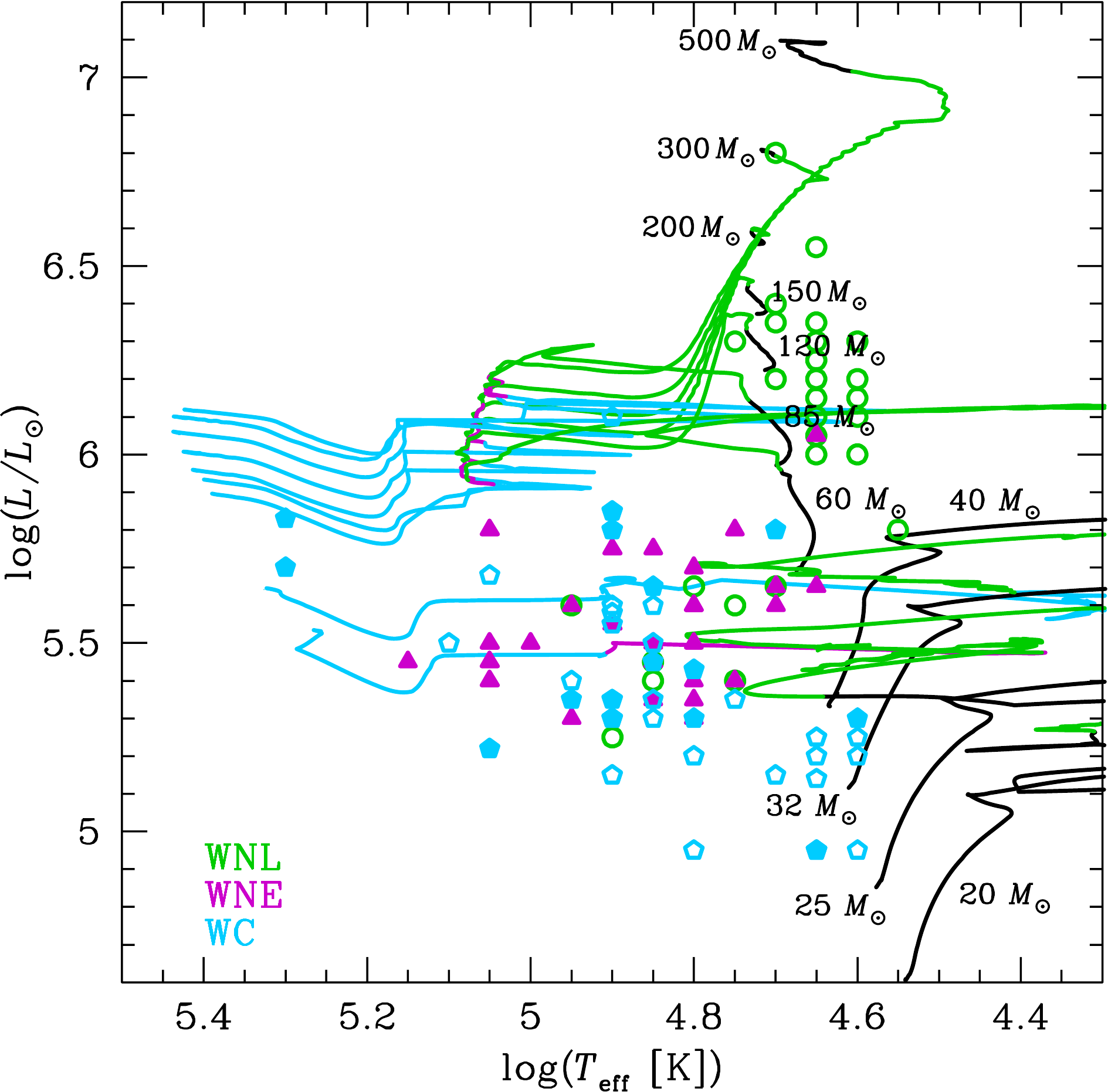}
\caption{The positions of WR stars observed by \citet{HGL06}
and \citet{SHT12} are indicated with the rotating evolutionary tracks taken from \citet{SE12} for masses up to 120 $M_\odot$ and from \citet{Liza13} for VMS.
}
\label{fig:wrobs}
\end{figure}

We see in Fig.~\ref{fig:wrobs} that VMS models well fit the most luminous WNL stars. On the other hand, they
predict very luminous WC stars. Of course the fact that no such luminous WC stars has ever been observed can simply come from the fact that such stars are very rare and the lifetime in the WC phase is moreover relatively short.

\section{Late evolution and pre-SN properties of Very Massive Stars}\label{fate}

The next chapter discusses the explosion that will take place at the end of VMS life but {whether or not a star produces a pair-instability supernova (PISN, aka pair-creation SN, PCSN)} can be reasonably {estimated} from the mass of its carbon-oxygen (CO) core as demonstrated by the similar fate for stars with the same CO core found in various studies of VMS in the early Universe \citep{BAC84,HEGER02,CE12,DWL13}, even if their prior evolution is different. In this section, we will thus use the CO core mass {to estimate} the fate of the models discussed in the previous sections\footnote{Note that for lower-mass massive stars ($\lesssim 50\,M_\odot$), the CO core mass alone is not sufficient to predict the fate of the star and other factors like compactness, rotation and the central carbon abundance at the end of helium burning also play a role \citep[see e.\,g.][]{CL13}.}. We will only briefly discuss the supernova types that these VMS may produce in this chapter as this is discussed in chapters 7 and 8.

\begin{table*}
\caption{Initial masses, mass content of helium in the envelope, mass of carbon-oxygen core, final mass in solar masses and fate of the models estimated from the CO core mass.}\label{table:fate}
\centering
\begin{tabular}{ccccc|ccccc}
\hline
&\multicolumn{4}{c|}{non-rotating}  &\multicolumn{4}{c}{rotating}\\
$M_{ini}$  &$M_{\textrm{He}}^{\textrm{env}}$  &$M_{\textrm{co}}$ &$M_\textrm{final}$ &Fate   &$M_{\textrm{He}}^{\textrm{env}}$  &$M_{\textrm{co}}$ &$M_\textrm{final}$ &Fate\\
\hline
\multicolumn{9}{c}{{Z=0.014}}  \\
120  &0.4874  &25.478 &30.8 &CCSN/BH   &0.5147  &18.414  &18.7 &CCSN/BH\\
150  &0.6142  &35.047 &41.2 &CCSN/BH   &0.5053  &19.942  &20.2 &CCSN/BH\\
200  &0.7765  &42.781 &49.3 &CCSN/BH   &0.5101  &21.601  &21.9 &CCSN/BH\\
300  &0.3467  &32.204 &38.2 &CCSN/BH   &0.4974  &19.468  &23.9 &CCSN/BH\\
500  &0.3119  &24.380 &29.8 &CCSN/BH   &0.5675  &20.993  &25.8 &CCSN/BH\\
\\
\multicolumn{9}{c}{{Z=0.006}}  \\
120  &1.2289  &43.851 &54.2 &CCSN/BH   &0.5665  &32.669  &39.2 &CCSN/BH\\
150  &1.1041  &47.562 &59.7 &CCSN/BH   &0.7845  &38.436  &45.6 &CCSN/BH\\
200  &-       &-      &-    &CCSN/BH   &0.5055  &42.357  &51.0 &CCSN/BH\\
300  &-       &-      &-    &CCSN/BH   &0.5802  &44.959  &54.0 &CCSN/BH\\
500  &1.6428  &92.547 &94.7 &PISN      &0.7865  &73.145  &74.8 &PISN\\
\\
\multicolumn{9}{c}{{Z=0.002}}  \\
150  &-  &-  &-     &-      &2.3353  &93.468  &106.5 &PISN\\
200  &-  &-  &-     &-      &3.3022  &124.329  &129.2 &PISN\\
300  &-  &-  &-     &-      &5.5018  &134.869  &149.7 &BH\\

\hline
\end{tabular}
\end{table*}

\subsection{Advanced phases, final masses and masses of carbon-oxygen cores}\label{tcrhoc}

\begin{figure*}
\centering
\begin{tabular}{cc}
\includegraphics[width=0.5\textwidth,clip=]{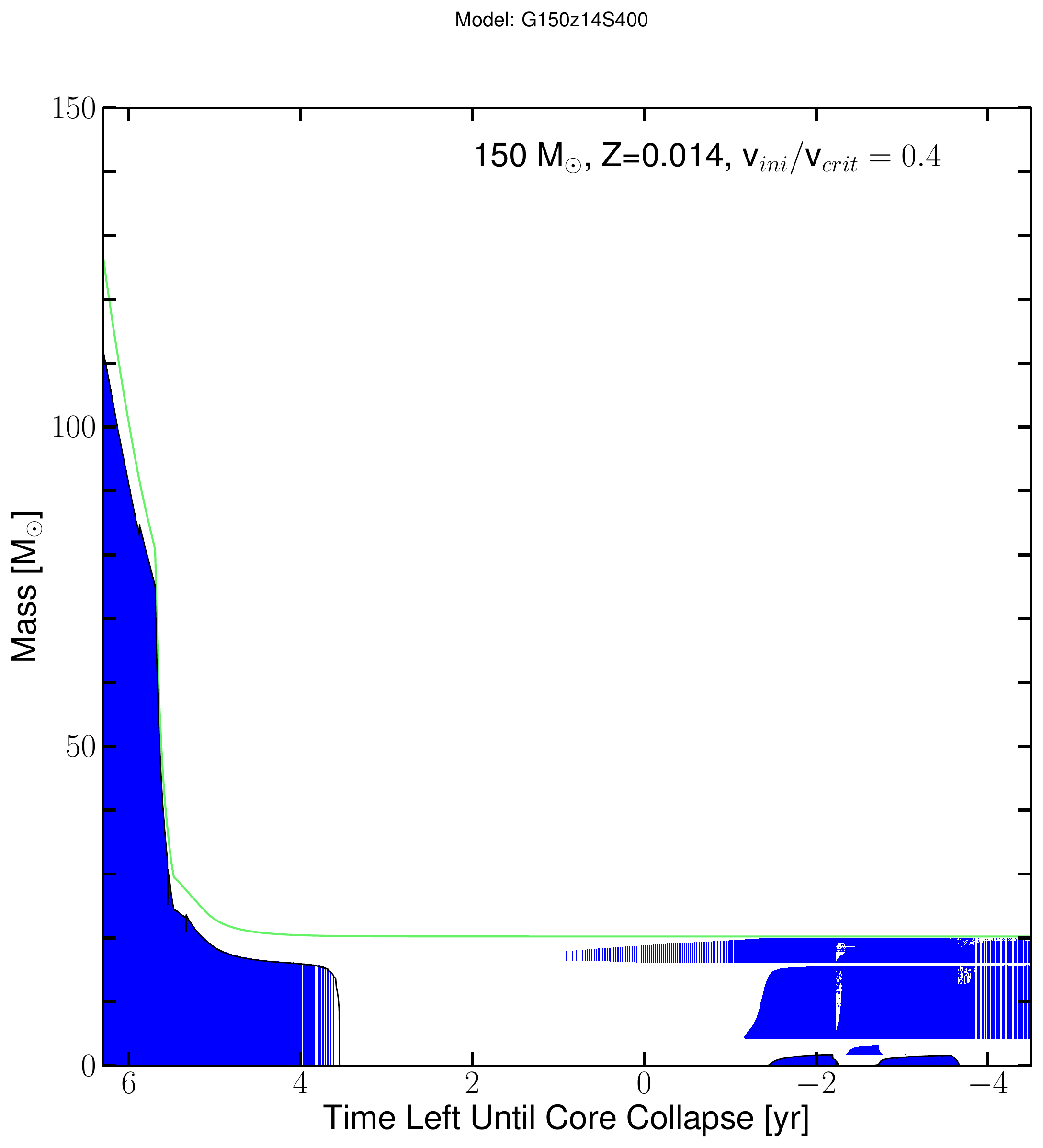} &
\includegraphics[width=0.5\textwidth,clip=]{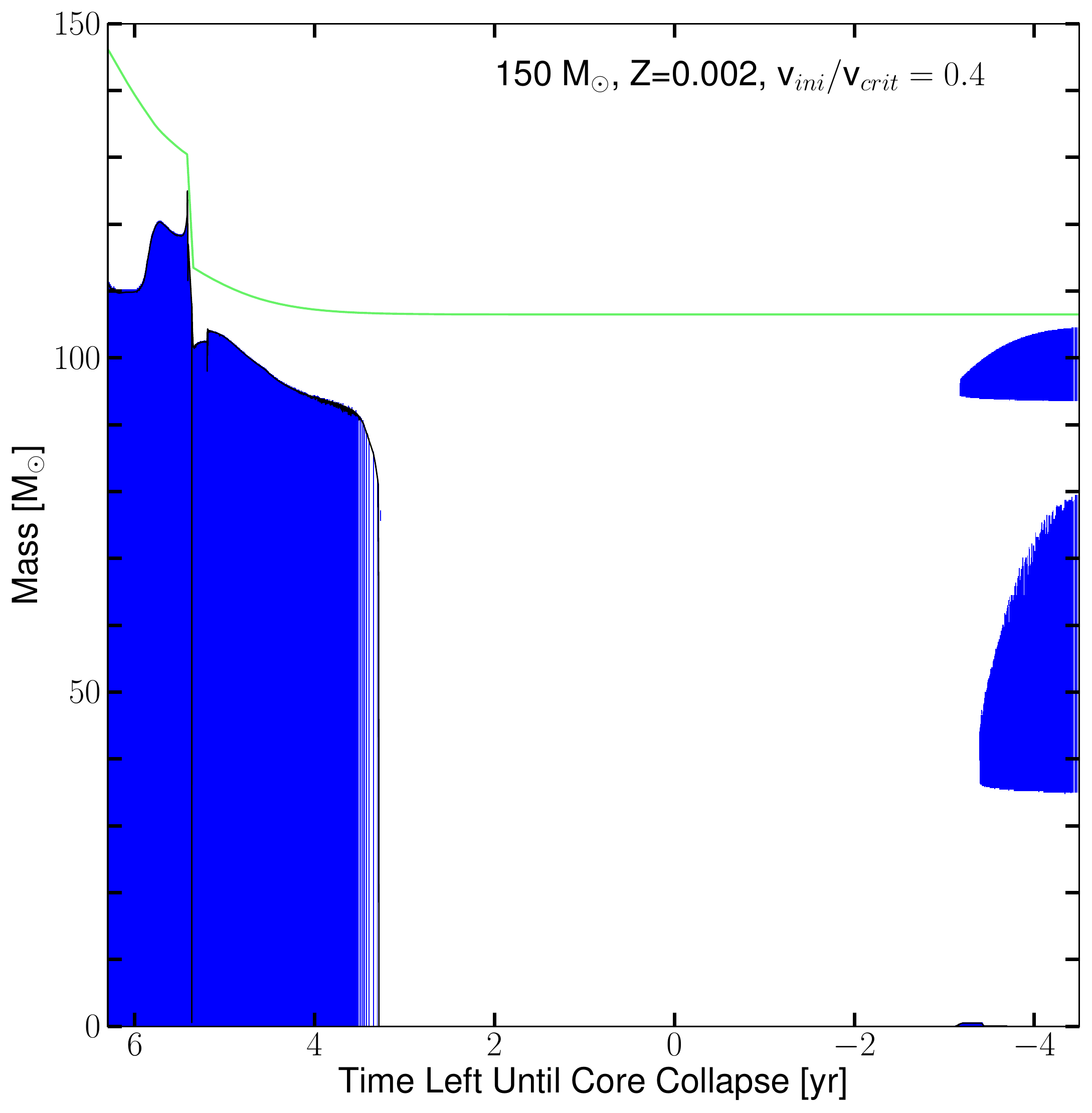} 
\end{tabular}
\caption{Structure evolution diagram for rotating 150 $M_\odot$ at solar and SMC metallicities as a function of the log of the time left until the last model. The blue zones represent the convective regions and the top solid line the total mass. }\label{fig:kip_logt}
\end{figure*}

In Fig. \ref{fig:kip_logt}, the structure evolution diagrams are drawn as a function of the log of the time left until the last model calculated (as opposed to age in Fig.\,\ref{fig:kip_age}). This choice of $x$-axis allows one to see the evolution of the structure during the advanced stages. In the {\it left} panel, we can see that, at solar metallicity, VMS have an advanced evolution identical to lower mass stars \citep[see e.\,g. Fig. 12 in][]{psn04} with a radiative core C-burning followed by a large convective C-burning shell, radiative neon burning and convective oxygen and silicon burning stages. All the solar metallicity models will eventually undergo core collapse after going through the usual advanced burning stages. {As presented in Table \ref{Table:endHe} (column 9), the central mass fraction of $^{12}$C is very low in all VMS models and is anti-correlated with the total mass at the end of helium burning (column 6): the higher the total mass, the lower the central $^{12}$C mass fraction. This is due to the higher temperature in more massive cores leading to a more efficient $^{12}$C($\alpha, \gamma$)$^{16}$O relative to 3$\alpha$.}

\begin{figure}
\center
\begin{tabular}{ccc}
\includegraphics[width=0.8\textwidth,clip=]{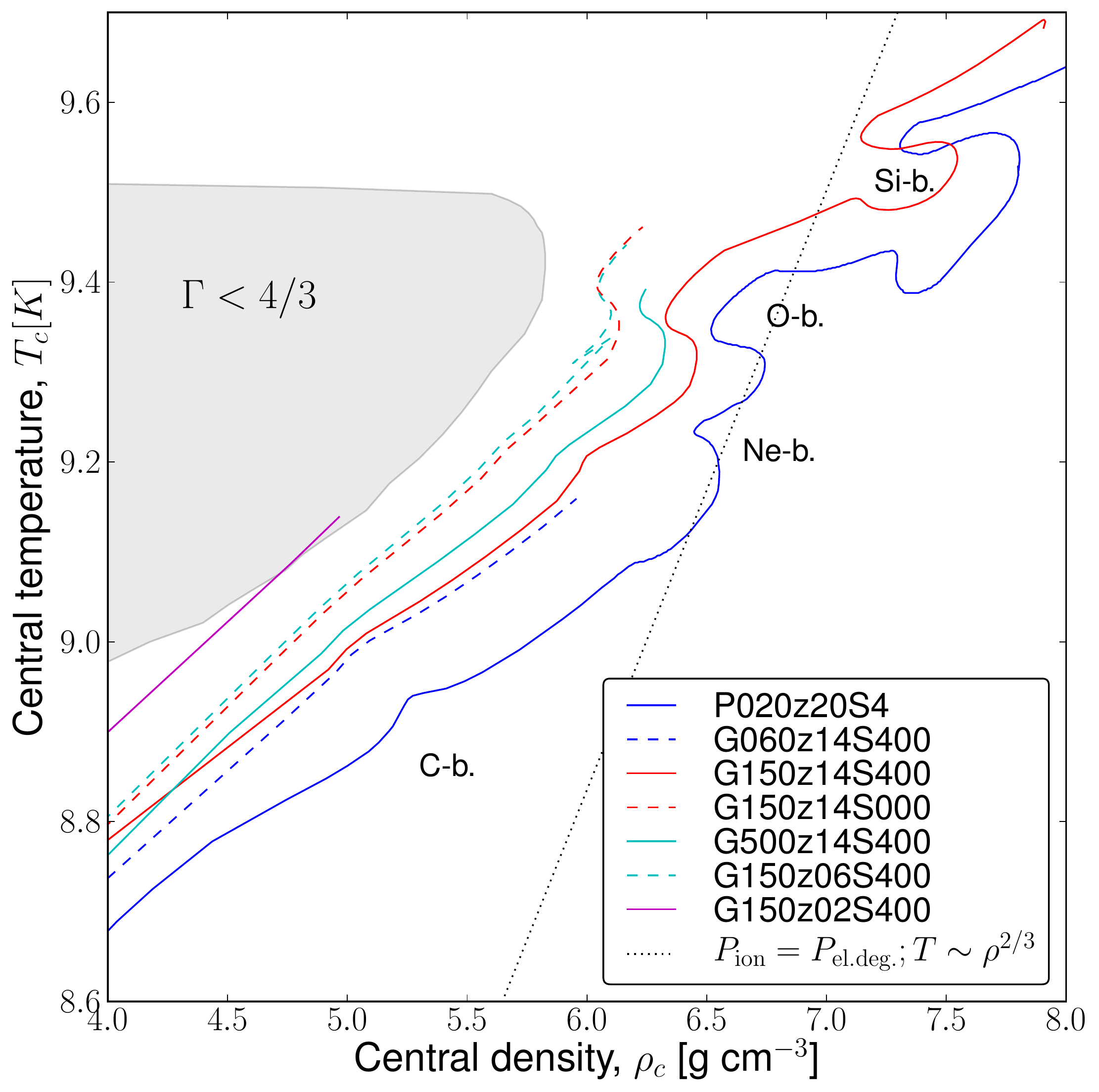} 
\end{tabular}
\caption{Evolution of the central temperature $T_c$ versus central density $\rho_c$ 
for the rotating 20 \citep[from][]{psn04}, 60 \citep[from][]{SE12}, 150 and 500 $M_\odot$ models and non-rotating 150 $M_\odot$ model at solar metallicity as well as the rotating 150 $M_\odot$ model at SMC 
metallicity. The gray shaded area is the pair-creation instability region 
($\Gamma < 4/3$, where $\Gamma$ is the adiabatic index). The additional dotted line corresponds to the limit 
between non-degenerate and degenerate electron gas.
} \label{fig:tcrhoc}
\end{figure}

The similarities between VMS and lower mass stars {at solar metallicity} during the advanced stages can also be seen in the central temperature versus central density diagram (see Fig.~\ref{fig:tcrhoc}). Even the evolution of the 500 $M_\odot$ rotating model is close to that of the 60 $M_\odot$ model. The non-rotating models lose less mass as described above and thus their evolutionary track is higher (see e.\,g. the track for the non-rotating 150 $M_\odot$ model in Fig.\,\ref{fig:tcrhoc}). Non-rotating models nevertheless stay clear of the pair-instability region ($\Gamma < 4/3$, where $\Gamma$ is the adiabatic index) in the centre.

The situation is quite different at SMC metallicity (see Fig. \ref{fig:kip_logt}, {\it right} panel). Mass loss is weaker and thus the CO core is very large (93.5 $M_\odot$ for this 150 $M_\odot$ model). Such a large core starts the advanced stages in a similar way: radiative core C-burning followed by a large convective C-burning shell and radiative neon burning. The evolution starts to diverge from this point onwards. As can be seen in $T_c$ vs $\rho_c$ plot, the SMC 150 $M_\odot$ model enters the pair-instability region. These models will thus have a different final fate than those at solar metallicity (see below).

Figure~\ref{fig:mfinal} (see also Table \ref{mdotrates}) shows the final masses of VMS as a function of the initial masses.
All models at solar $Z$, rotating or not, end with a small fraction of their initial mass due to the strong mass loss they experience.
Rotation enhances mass loss by allowing the star to enter the WR phase earlier during the MS (see {\it top} panels of Fig.~\ref{fig:kip_age}) and the final mass of non-rotating models is generally higher than that of rotating models. 
At low metallicities, due to the metallicity dependence of radiatively-driven stellar winds in both O-type stars \citep{VN01} and WR stars \citep{EV06}, final masses are larger.

\begin{figure}
\includegraphics[width=0.8\textwidth]{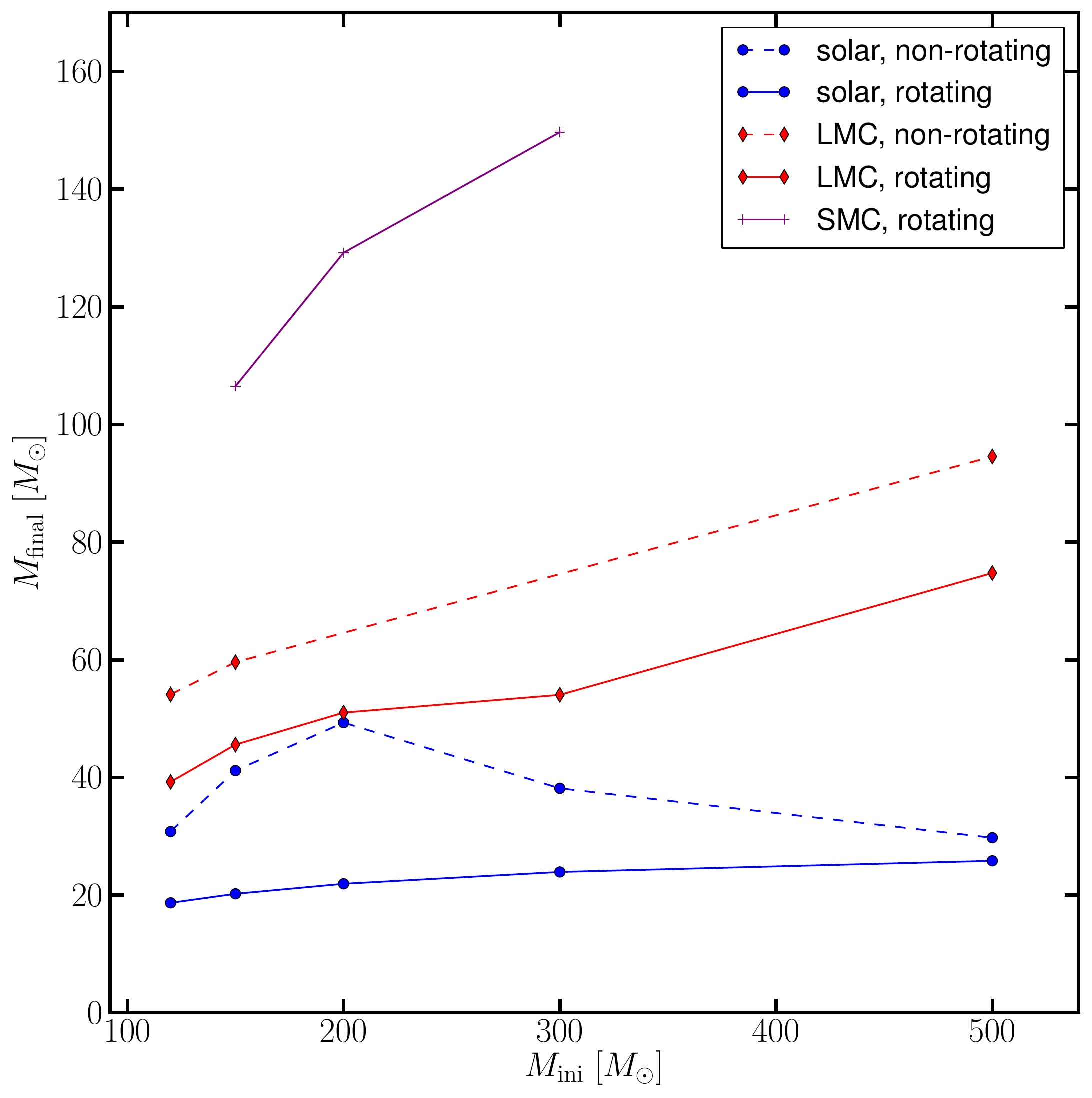}
\caption{Final mass versus initial mass for all rotating (solid lines) and non-rotating (dashed line) models.} \label{fig:mfinal}
\end{figure}

\begin{figure}
\includegraphics[width=0.8\textwidth]{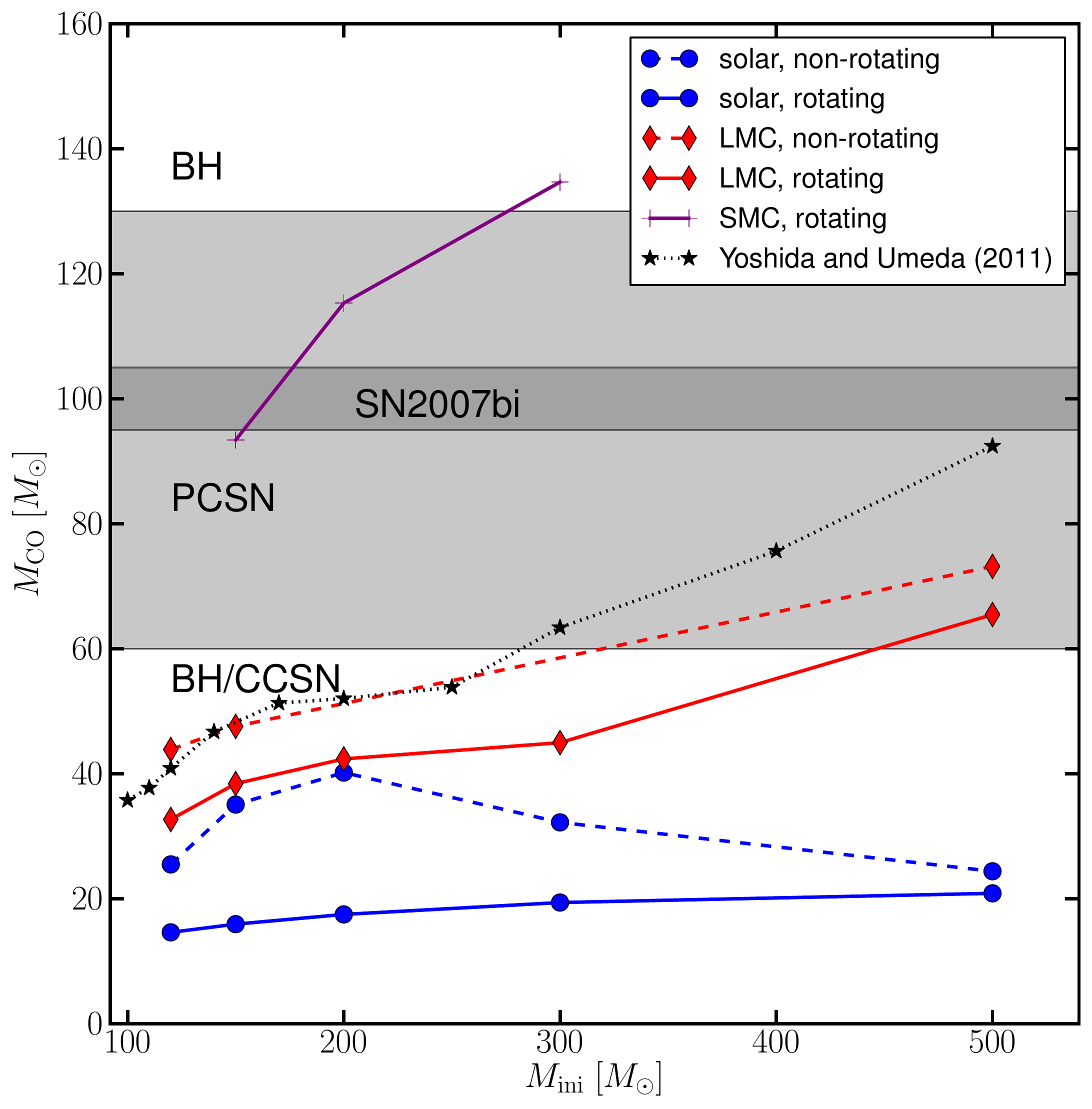}
\caption{Mass of carbon-oxygen core of all the models as a function of the initial mass. The light grey shaded area represents the range of $M_{CO}$, for which the estimated fate is a PISN. The thin dark grey shaded area corresponds to the estimated $M_{CO}$ of the progenitor of SN2007bi assuming it is a PISN (see text for more details). The points linked by the dotted black line are from the models of \citet{YH11} at $Z=0.004$, case A.  } \label{fig:mco}
\end{figure}

Fig. \ref{fig:mco} shows how the CO core masses vary as a function of the initial mass, rotation and metallicity.  The CO core ($M_{\textrm{CO}}$) is here defined as the core mass for which the mass fraction of C+O is greater than 75\%.
Since the CO core mass is so close to the total mass, the behavior is the same as for the total mass and for the same reasons. For the rotating solar metallicity models, mass loss is so strong that all models end with roughly the same CO core mass around 20 $M_\odot$. As the metallicity decreases, so does mass loss and thus the LMC and SMC models have higher final CO core masses and the CO core mass does depend on the initial mass in a monotonous way. Finally, non-rotating models lose less mass than their rotating counterpart since they enter the WR phase later and also have less hot surface.
Simulations at $Z=0.004$ from \citet{YH11} (case A) are also plotted in Fig. \ref{fig:mco}. The CO core masses they obtain are consistently slightly larger than for the LMC ($Z=0.006$) models. 

\subsection{Do VMS produce PISNe?}\label{pcsn}

As mentioned above, the core masses, especially the CO core masses, can be used to estimate whether or not models produce a Pair Instability SuperNova (PISN) by using the results of previous studies, which follow the explosion of such massive cores and knowing that VMS with the same CO core masses have similar core evolution from carbon burning onwards. \citet{HEGER02} calculated a grid of models and found that stars with helium cores ($M_{\alpha}$) between 64 and 133 $M_\odot$ produce PISNe and that stars with more massive $M_{\alpha}$ will collapse to a BH without explosion, confirming the results of previous studies, such as \citet{BAC84}. {The independent results of \citet{CE12} also confirm the CO core mass range that produce PISNe.}

PISNe occur when very massive stars (VMS) experience an instability in
their core during the neon/oxygen burning stage due to the creation of electron-positron
pairs out of two photons. The creation of pairs in their oxygen-rich core softens the equation of
state, leading to further contraction. This runaway collapse
is predicted to produce a very powerful explosion, in excess of $10^{53}$ erg, disrupting the entire star and leaving no remnant \citep{BAC84, FCL01}.

 \citet{HEGER02} also find that stars with $M_{\alpha}$ between roughly 40 and 63 $M_\odot$ will undergo violent pulsations induced by the pair-instability leading to strong mass loss but which will not be sufficient to disrupt the core. Thus these stars will eventually undergo core collapse as lower mass stars. Since in our models, the CO core masses are very close to $M_{\alpha}$ (equal to the final total mass in our models, see Table \ref{table:fate}), in this chapter we assume that models will produce a PISN if 60 $M_\odot\leq M_{CO}\leq$ 130 $M_\odot$. In Fig. \ref{fig:mco}, the light grey shaded region corresponds to the zone where one would expect a PISN, the dark shaded region show the
estimated range of the carbon oxygen core of the progenitor of SN2007bi \citep[see][for more details]{Liza13}.

We see in Fig. \ref{fig:mco} that at solar metallicity none of the models is expected to explode as a PISN. At the metallicity of the LMC, only stars with initial masses above 450 for the rotating models and
above about 300\,$M_\odot$ for the non-rotating case are expected to explode as a PISN.
At the SMC metallicity, the mass range for the PISN progenitors is much more favorable.
Extrapolating the points obtained from our models we obtain that all stars in the mass range
between about 100\,$M_\odot$ and 290\,$M_\odot$ could produce PISNe. Thus these models provide support for the occurrence of PISNe in the nearby (not so metal poor) universe.

Table \ref{table:fate} presents for each of the models, the initial mass ($M_{\rm ini}$), the amount of helium left in the star at the end of the calculation ($M_{\textrm{He}}^{\rm env}$),  and final total mass as well as the estimated fate in terms of the explosion type: PISN or core-collapse supernova and black hole formation with or without mass ejection (CCSN/BH). The helium core mass ($M_\alpha$) is not given since it is always equal to the final total mass, all the models having lost the entire hydrogen-rich layers.

\subsection{Supernova types produced by VMS}

Let us recall that, in VMS, convective cores are very large.
It is larger than 90\% above 200 $M_\odot$ at the start of the evolution and even though it decreases slightly during the evolution, at the end of core H-burning, the convective core occupies more than half of the initial mass in non-rotating models and most of the star in rotating models. This has an important implication concerning the type of supernovae that these VMS will produce. Indeed, even if mass loss is not very strong in SMC models, all the models calculated have lost the entire hydrogen rich layers long before the end of helium burning. 
Thus the models predict that all VMS stars in the metallicity range studied will produce either a type Ib or type Ic SN but no type II.
\subsection{GRBs from VMS?}\label{grb}

\begin{figure*}
\includegraphics[width=0.3\textwidth,clip=]{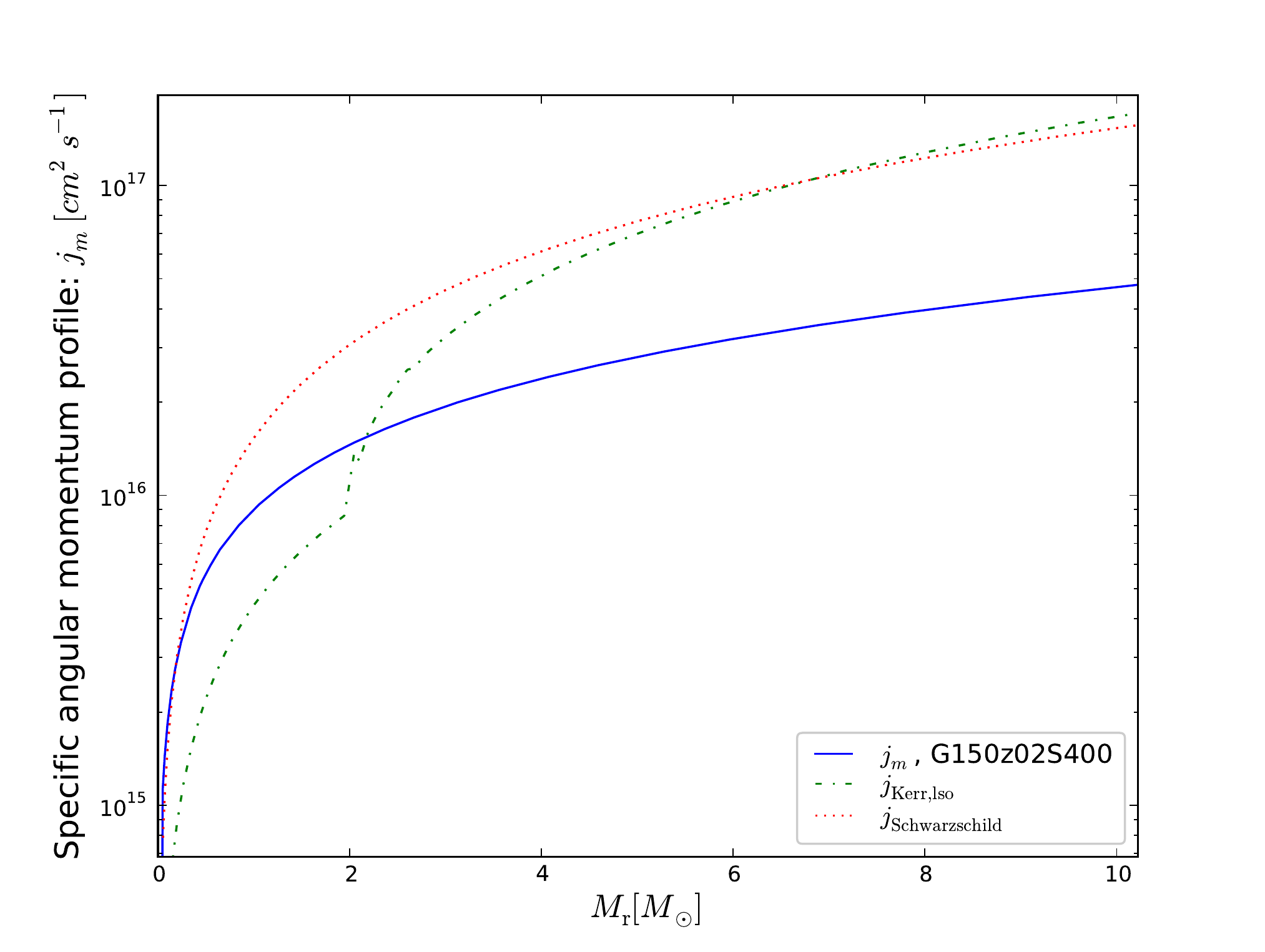}\includegraphics[width=0.3\textwidth,clip=]{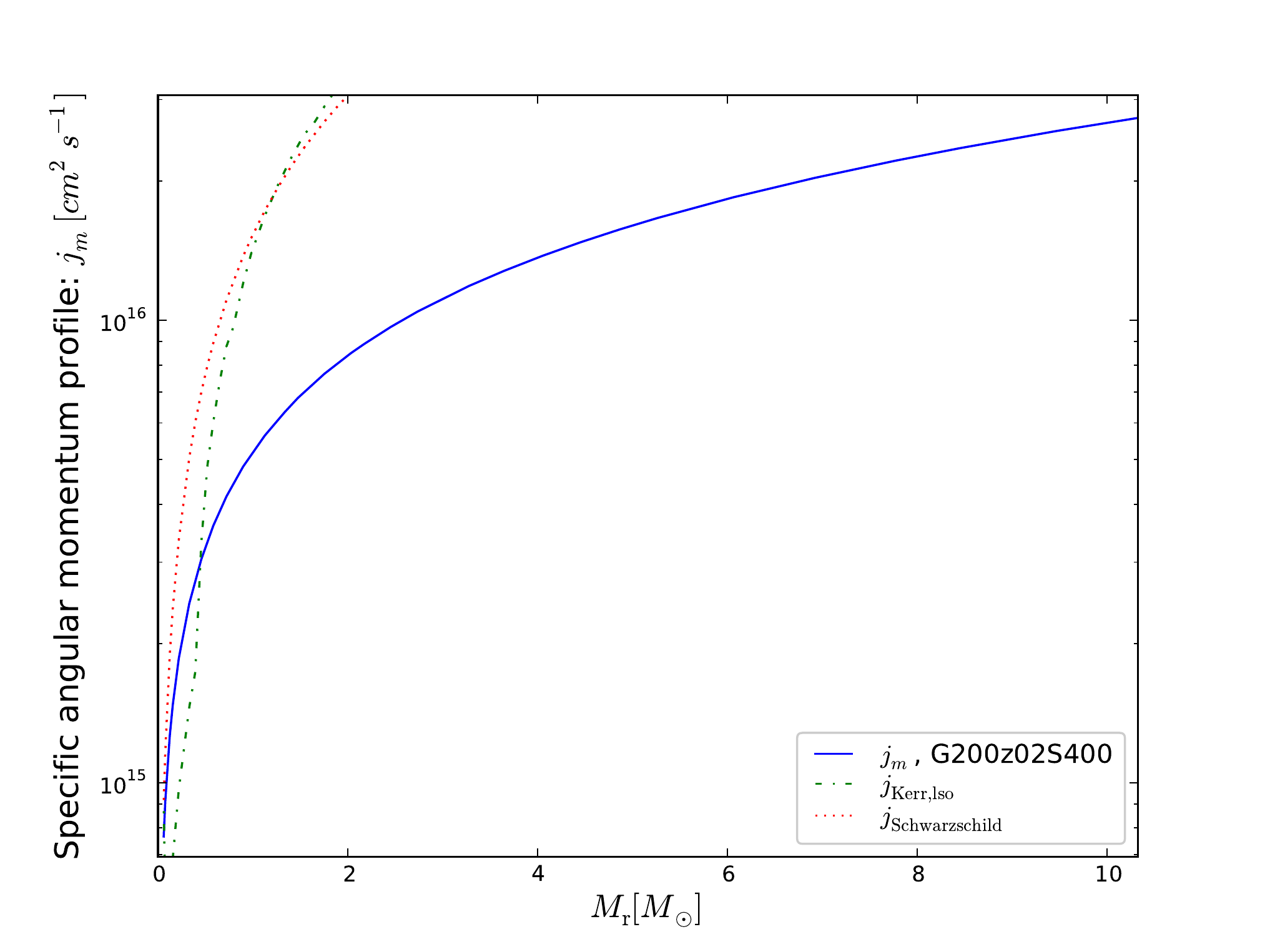}\includegraphics[width=0.3\textwidth,clip=]{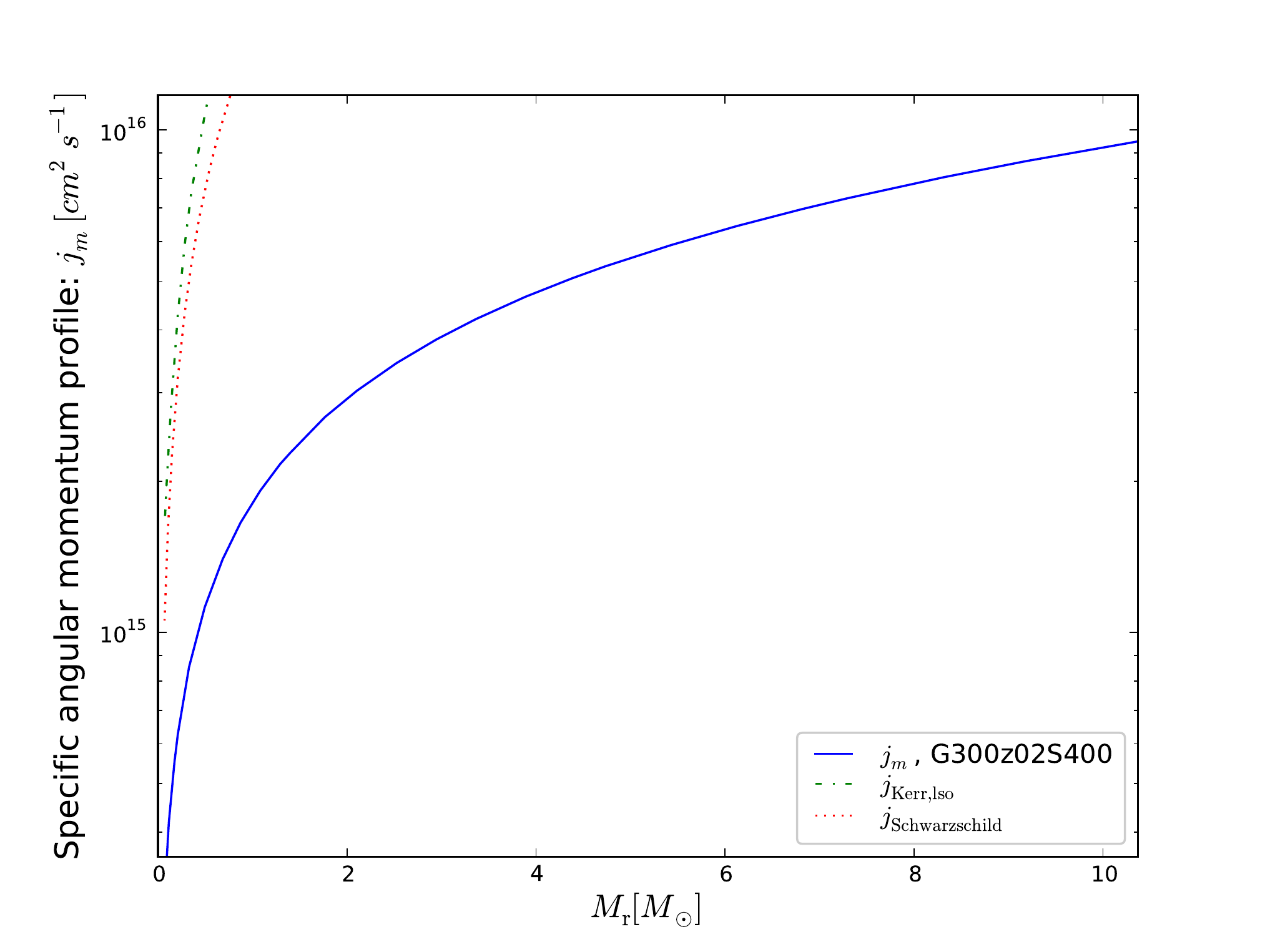}
\caption{Specific angular momentum profile, $j_\mathrm{m}$, as a function of the Lagrangian mass coordinate in the core of the SMC rotating 150, 200, 300 $M_\odot$ models, plotted at the end of the calculations (solid line). The dash-dotted line is $j_{\rm Kerr,lso}=r_{\rm LSO}\,c$ \citep[ p. 428]{ST83}, where the radius of the last stable orbit, $r_{\rm LSO}$, is given by $r_{\rm ms}$ in formula (12.7.24) from \citet[p. 362]{ST83} for circular orbit in the Kerr metric. $j_{\rm Kerr,lso}$ is the minimum specific angular momentum necessary to form an accretion disc around a rotating black hole. $j_\mathrm{Schwarzschild}=\sqrt{12}Gm/c$ (dotted line) is the minimum specific angular momentum necessary for a non-rotating black hole, for reference.} \label{fig:grb}
\end{figure*}

The evolution of the surface velocity was described in Sect. \ref{vsurf}. Only models at SMC retain a significant amount of rotation
during their evolution {(see angular momentum contained in the CO core at the end of helium burning in the last column of
Table \ref{Table:endHe})} but do they retain enough angular momentum for rotation to affect the fate of the star? The angular momentum profile of the SMC models is presented in Fig. \ref{fig:grb}. Note that the models presented in this section do not include the Tayler-Spruit dynamo so represent the most optimistic (highest possible) prediction concerning the angular momentum in the core of these models. Mass loss
in the 300\,$M_\odot$ model is too strong for the core to retain enough angular momentum for rotation to impact the death of this
model. In the 200\,$M_\odot$ model, and even more so in the 150\,$M_\odot$ model, however, the central part of the core retain a
significant amount of angular momentum that could potentially affect the death of the star. {Since the role of rotation is
very modest from carbon until just after the end of core silicon burning, even for extremely fast rotators \citep[see
e.\,g.][]{HMM05,CL13}, we do not expect rotation to affect significantly the fate of stars that are predicted to explode as PISN
during neon-oxygen burning. However, as discussed in \citet[][and references therein]{YDL12}, the large angular momentum content is
most interesting for the stars that just fall short of the minimum CO core mass for PISN \citep[since fast rotation plays an
important role during the early collapse][]{OTT04,CON11,CL13}.} Indeed, without rotation, these stars would produce a BH following a possible pulsation pair-creation phase, whereas with rotation, these stars could produce energetic asymmetric explosions (GRBs or magnetars). Since the 150 $M_\odot$ model is predicted to explode as a PISN, we thus do not expect the models presented in this grid to produce GRBs or magnetars but such energetic asymmetric explosions are likely to take place in lower mass and lower metallicity stars \citep[see][]{HMM05,YL05,WH06}.

\citet{YDL12} calculated a grid of zero-metallicity rotating stars, including the Tayler-Spruit dynamo for the interaction between rotation and magnetic fields. They find that fast rotating stars with an initial mass below about 200 $M_\odot$ retain enough angular momentum in their cores in order to produce a collapsar \citep[$j>j_{\textrm Kerr, lso}$][]{W93} or a magnetar \citep[see e.\,g.][]{W00,BDL07,DOO12}. Thus some VMS that do not produce PISNe might produce GRBs instead.

\section{The final chemical structure and contribution to galactic chemical evolution}

\begin{figure}
\center
\includegraphics[width=0.8\textwidth,clip=]{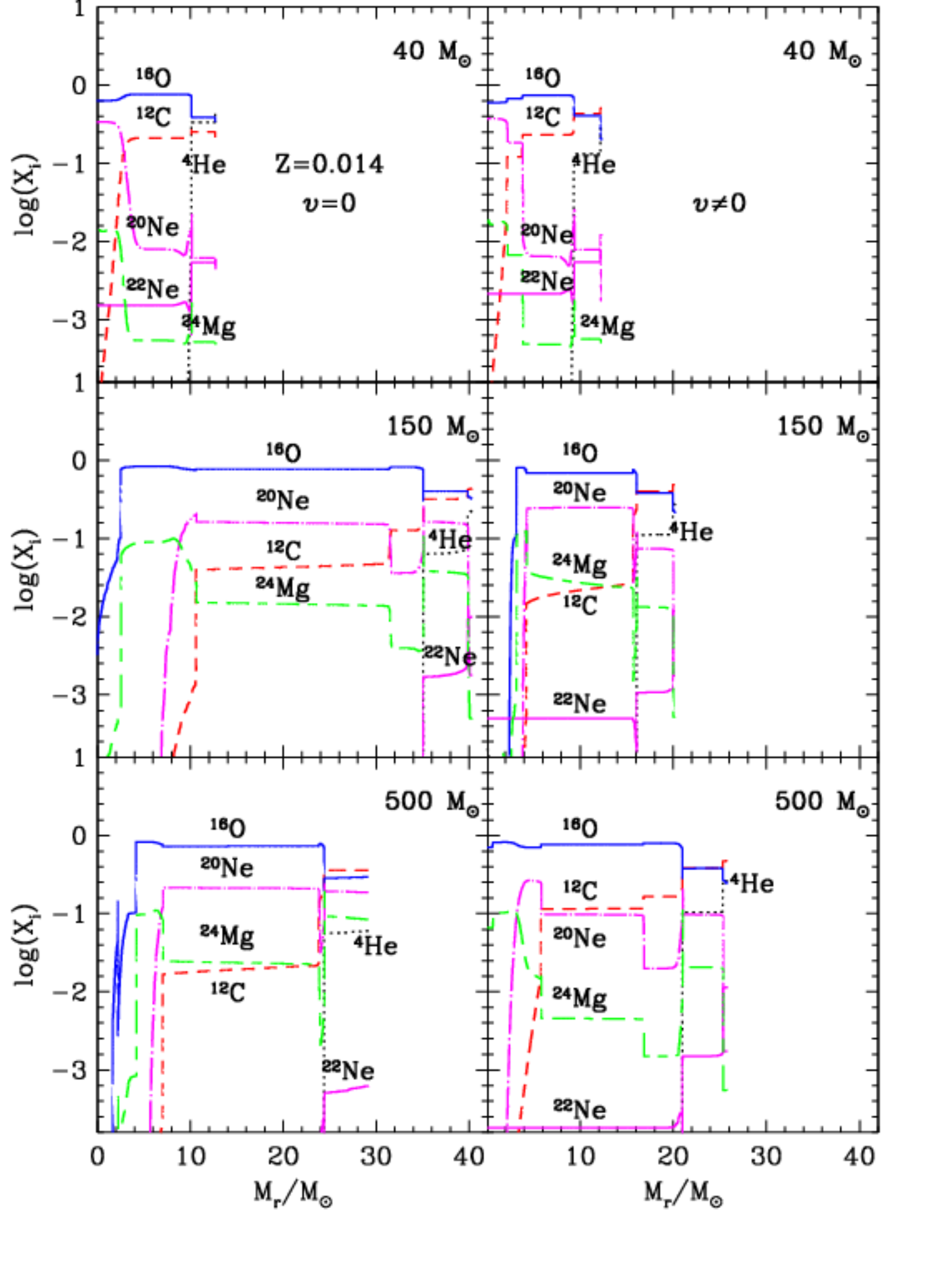}
\caption{Chemical structure of 40, 150 and 500 $M_\odot$ non-rotating ({\it left}) and rotating ({\it right}) models at Z=0.014 at the end of the calculations. Note that the rotating 500 M model is shown at an earlier evolutionary stage than the corresponding non-rotating model.
}
\label{fig:finalc}
\end{figure}

Figure \ref{fig:finalc} shows the chemical structure at the last time steps calculated, which is the end of the carbon burning phase in the case of the 40 $M_\odot$, and the end of the core oxygen-burning phase in the case of the 150 and 500 $M_\odot$ models.
A few interesting points come out from considering this figure. First, in all cases, some helium is still present in the outer layers. Depending on how the final stellar explosion occur, this helium may or may not be apparent in the spectrum, as discussed in \citet{Liza13}. 
Second, just below the He-burning shell, products of the core He-burning, not affected by further carbon burning are apparent. This zone extends between about 4 and 10 $M_\odot$ in the 40 $M_\odot$ model, between about 32 and 35 $M_\odot$ in the 150 $M_\odot$ model and in a tiny region centered around 24 $M_\odot$ in the 500 $M_\odot$ model. 
We therefore see that this zone decreases in importance when the initial mass increases. Interestingly, the chemical composition in this zone present striking differences if we compare for instance the 40 $M_\odot$ and the 500 $M_\odot$ model. We can see that the abundance of $^{20}$Ne is much higher in the more massive model. This comes from the fact that in more massive stars, due to higher central temperatures during the core He-burning phase the reaction $^{16}$O($\alpha$, $\gamma$)$^{20}$Ne is more active, building thus more $^{20}$Ne. Note that $^{24}$Mg is also more abundant, which is natural since the reaction $^{20}$Ne($\alpha$, $\gamma$)$^{24}$Mg reaction will also be somewhat active in VMS for the same reasons. While in the case of the 150 $M_\odot$, due to the mass loss history, the $^{20}$Ne and $^{24}$Mg-rich layers are not uncovered, they are uncovered in the 500 $M_\odot$ model. This implies that strong overabundances of these two isotopes at the surface of WC stars can be taken as a signature for an initially very massive stars as the progenitor of that WC star. It also means that, contrary to what occurs at the surface of WC stars originating from lower initial mass stars, neon is no longer present mainly in the form of $^{22}$Ne (and thus be a measure of the initial CNO content since resulting from the transformation of nitrogen produced by CNO burning during the H-burning phase) but will mainly be present in the form of $^{20}$Ne.
 
Rotation does not change much this picture (see right panel of Fig.~\ref{fig:finalc}), except that, due to different mass loss histories,
the rotating models lose much more mass and end their evolution with smaller cores. This is particularly striking for the 150 $M_\odot$ model.
Qualitatively the situation is not much different at lower metallicities.

As stars in the mass range 50-100\,$M_\odot$ \citep[see e.\,g.][]{ROTXI,CL13}, VMS eject copious amount of H-burning products through their stellar winds and to a lesser extent He-burning products. The potential difference between VMS and stars in the mass range 50-100\,$M_\odot$ is how they explode (or not) at the end of their life, which is discussed in detail in the next chapter. If they collapse to a black hole the contribution from the supernova will be negligible whereas if they explode as PISNe they will produce large amounts of iron and other heavy elements. 

Do VMS stars contribute to the chemical enrichment of galaxies or are VMS so rare that whatever their evolution, their impact on energy and mass outputs will anyway be very low?
Considering a Salpeter IMF, the number of stars with masses 
between 120 and 500  $M_{\odot}$ corresponds to only about 2\% of the total number of stars with masses between 8 and 500 $M_{\odot}$. 
So they are indeed only very few!
On the other hand, one explosion can release a great amount of energy and mass into the interstellar medium. 
Typically a 200  $M_{\odot}$ star releases about ten times more mass than a 20 $M_{\odot}$ star. 
If we roughly suppose that for hundred 20 $M_{\odot}$ stars there are only two 200 $M_{\odot}$ star, this means that the 200 $M_{\odot}$ stars contribute to the release of mass at a level corresponding to about 20\% of the release of mass by 20 $M_{\odot}$, which is by far not negligible. Of course this is a rough estimate but, as a rule of thumb we can say that any quantity released by a VMS $\sim$ tenfold
intensity compared to that of a typical, 20 $M_{\odot}$ star will make a
non-negligible difference in the overall budget of this quantity at the level of a galaxy. 
For instance, the high bolometric
luminosities, stellar temperatures and mass loss rates of VMS imply
that they will contribute significantly to the radiative and mechanical
feedback from stars in high mass clusters at ages prior to the first 
supernovae \citep{DVLT13,PAC10}. Core-collapse SNe produce of the order of $0.05\,M_\odot$ (ejected masses) of iron, $1\,M_\odot$ of each of the $\alpha-$elements. According to the production factors in Table 4 in \citet{HEGER02}, PISN produce up to $40\,M_\odot$ of iron, of the order of $30\,M_\odot$ of oxygen and silicon and of the order of $5-10\,M_\odot$ of the other $\alpha-$elements. Considering that PISN may occur up to SMC metallicity and represent 2\% of SNe at a given metallicity, their contribution to the chemical enrichment of galaxies may be significant, especially in the case of iron, oxygen and silicon. If the IMF is top heavy at low metallicities, the impact of VMSs would be even larger.

\section{Summary and Conclusion}

In this chaper, we have discussed the evolution of very massive stars based on stellar evolution models at various metallicities. The main properties of VMS are the following:
\begin{itemize}
\item VMS possess very large convective cores during the MS phase. Typically, in a 200 $M_\odot$ model on the ZAMS
the convective core extends  over more than 90\% of the total mass.
\item Since the mass-luminosity relation flattens above 20 $M_\odot$, VMS have lifetimes that are not very sensitive to their initial mass and range between 2 and 3.5 million years. 
\item Even in models with no rotation, due to the importance of the convective core, VMS stars evolve nearly chemically homogeneously.
\item Most of the very massive stars (all at solar $Z$) remain in blue regions of the HR diagram and do not go through a luminous blue variable phase.
\item They all enter into the WR phase and their typical evolution is Of - WNL-  WNE - WC/WO.
\item Due to increasing mass loss rates with the mass, very different initial mass stars end with similar final masses. As a consequence very different initial masses may during some of their evolutionary phases occupy very similar positions in the HRD.
\item A significant proportion of the total stellar lifetimes of VMS is spent in the WR phase (about a third).
\item A WC star with high Ne ($^{20}$Ne) and Mg  ($^{24}$Mg) abundances at the surface has necessarily a VMS as progenitor.
\item At solar metallicity VMS are not expected to explode as PISNe because mass loss rates are too high. 
\item Whether or not some VMS models retain enough mass to produce a PISN at low metallicity is strongly dependent on mass loss. As discussed above, models that retain enough mass at SMC metallicity (and below) also approach very closely the Eddington limit after helium burning and this might trigger a strong enough mass loss in order to prevent any VMS from producing a PISN.
\item Most VMS lose the entire hydrogen rich layers long before the end of helium burning. 
Thus most VMS stars near solar metallicity are expected to produce either a type Ib or type Ic SN but no type II.
\item Models near solar metallicity are not expected to produce GRBs or magnetars. The reason for that is that either they lose too much angular momentum by mass loss or they avoid the formation of a neutron star or BH because they explode as PISN. Lower mass stars at low metallicities ($Z\lesssim 0.002$), however, may retain enough angular momentum as in metal free stars \citep[see][]{YDL12,CE12} for rotation (and magnetic fields) to play a significant role in their explosion.
\end{itemize}

Even though VMS are rare, their extreme luminosities and mass loss will still contribute significantly to the light and chemistry budget of their host galaxies. And although many VMS will die quietly and form a black hole, some VMS may die as PISNe or GRBs. Thus the extreme properties of VMS compensate for their rarity and they are worth studying and considering in stellar population and galactic chemical evolution studies. As discussed above, the main uncertainty that strongly affects their evolution and their fate is the uncertainty in mass loss, especially for stars near the Eddington limit. Thus, although the discussions and conclusions presented in this chapter will remain qualitatively valid, quantitative results will change as our knowledge of mass loss in these extreme stars improves.

\begin{acknowledgement}

The author thanks his collaborators at the University of Keele (C. Georgy), Geneva (G. Meynet, A. Maeder and Sylvia Ekstr\"om) and Malaysia (N. Yusof and H. Kassim) for their significant contributions to the results presented in this chapter. 
R. Hirschi acknowledges support from the World Premier International Research
Center Initiative (WPI Initiative), MEXT, Japan and from the Eurogenesis EUROCORE programme. The research leading to these results has received funding from the European Research Council under the European Union's Seventh Framework Programme (FP/2007-2013) / ERC Grant Agreement n. 306901. 
\end{acknowledgement}
\bibliographystyle{aa}

\end{document}